\documentclass[twocolumn,prl]{revtex4-1}
\usepackage[latin9]{inputenc}
\usepackage{xcolor}
\usepackage{amsmath}
\usepackage{amssymb}
\usepackage{graphicx}

\makeatletter

\usepackage{bm}

\usepackage[]{sidecap}
\sidecaptionvpos{figure}{c}

\makeatother

\begin{document}
\title{Ballistic charge transport in twisted bilayer graphene}
\author{Hadi Z. Olyaei$^{1}$, Bruno Amorim$^{2}$, Pedro Ribeiro$^{1,3}$,
Eduardo V. Castro$^{3,4}$}
\affiliation{$^{1}$CeFEMA, Instituto Superior Técnico, Universidade de Lisboa,
Av. Rovisco Pais, 1049-001 Lisboa, Portugal}
\affiliation{$^{2}$Centro de Física das Universidades do Minho e Porto, Universidade
do Minho, Campus de Gualtar, 4710-057 Braga, Portugal}
\affiliation{$^{3}$Beijing Computational Science Research Center, Beijing 100084,
China}
\affiliation{$^{4}$Centro de F\'{\i}sica das Universidades do Minho e Porto, Departamento
de F\'{\i}sica e Astronomia, Faculdade de Ciências, Universidade do
Porto, 4169-007 Porto, Portugal}
\begin{abstract}
We study conductance across a twisted bilayer graphene coupled to
single-layer graphene leads in two setups: a flake of graphene on
top of an infinite graphene ribbon and two overlapping semi-infinite
graphene ribbons. We find conductance strongly depends on the angle
between the two graphene layers and identify three qualitatively different
regimes. For large angles ($\theta\gtrsim10^{\circ}$) there are strong
commensurability effects - for incommensurate angles the low energy
conductance approaches that of two disconnected layers, while sharp
conductance features correlate with commensurate angles with small
unit cells . For intermediate angles ($3^{\circ}\lesssim\theta\lesssim10^{\circ}$),
we find a one-to-one correspondence between certain conductance features
and the twist-dependent Van Hove singularities arising at low energies,
suggesting conductance measurements can be used to determine the twist
angle. For small twist angles ($1^{\circ}\lesssim\theta\lesssim3^{\circ}$),
commensurate effects seem to be washed out and the conductance becomes
a smooth function of the angle. In this regime, conductance can be
used to probe the narrow bands, with vanishing conductance regions
corresponding to spectral gaps in the density of states, in agreement
with recent experimental findings.

\end{abstract}
\maketitle

\section{Introduction}

Two-dimensional materials have shown to be widely tunable, providing
properties-on-demand for electronic and optical applications. At the
same time, their relative simplicity and the degree of sample-purity
render these materials an ideal playground to study physical phenomena
hard to isolate in more complex compounds. The discovery of both correlated
insulating phases \citep{Cao2018a} and superconductivity \citep{Cao2018}
in twisted bilayer graphene (tBLG) opened up the possibility to further
tune 2D heterostructures into strongly correlated phases and promises
to shine new light on the interplay between the insulating and superconducting
states.

A key underlying feature of tBLG is the sharp decrease of the Fermi
velocity, strongly renormalized for small values of the twist angle,
$\theta$, between the two graphene layers \citep{dSPN}. For $\theta\lesssim1^{\circ}$,
extremely narrow bands appear at low energies, with the Fermi velocity
even vanishing at specific - so-called \emph{magic - }angles \citep{bistritzerPNAS2011,LopesDosSantos2012,TramblydeLaissardiere2010a,SuarezMorell2010,Shallcross2010}.
It is in this regime that unexpected insulating and superconducting
states are observed, pointing to electron correlations as key players.
This triggered an intense research interest in this system \citep{Chung2018,Yankowitz2019,Sharpe2019,Kerelsky2018,Choi2019,Jiang2019,Xie2019,Tomarken2019,Lu2019,Codecido2019,Shi2019,Serlin2019}.
The flat band regime induced by a finite twist has also been explored
experimentally in graphene double bilayers \citep{Shen2019,Liu2019a,Cao2019}
and trilayers \citep{Chen2019,Zuo2018}, and is also relevant to other
two-dimensional materials \citep{Wu2019,Wang2019}.

The rotation between layers introduces a long-wavelength modulation
of the lattice structure called moiré pattern \citep{Goncalo2019}.
For small angles, the moiré wavelength is much larger than the carbon-carbon
distance, with the ratio growing as $1/\theta$. This represents an
additional difficulty regarding the theoretical description of the
system, since a single moiré may contain several thousands of atoms
in the low angle regime. While the minimum Wannier-like tight binding
parametrization describing the narrow band sector on the moiré scale
is still debated \citep{Yuan2018a,Zou2018,Kang2018,Po2018a,Yuan2018,Carr2019a},
the original atomic tight binding model remains a faithful description,
despite the large unit cells at smaller angles \citep{bistritzerPNAS2011,LopesDosSantos2012,SuarezMorell2010,Shallcross2010,Shallcross2013,Sboychakov2015}.

Transport measurements in tBLG have been crucial to characterize the
small angle regime around the neutrality point. \emph{Superlattice}-induced
transport gaps, delimiting so-called moiré bands, have been observed
in Refs.~\citep{jarilloHerrero2016,Kim2016,Chung2018} for tBLG with
$\theta\sim2^{\circ}$, whenever these bands are completely occupied,
for a carrier density $n=+n_{\mathtt{S}}$, or fully empty, $n=-n_{\mathtt{S}}$,
where $2n_{\mathtt{S}}$ is the total charge density of the moiré
bands. Close to the first magic angle, for $\theta\approx1.1^{\circ}$,
besides the gaps at $n=\pm n_{\mathtt{S}}$, an insulating state was
also detected at half-filling, $n=\pm n_{\mathtt{S}}/2$, which cannot
be explained within the single particle picture \citep{Cao2018a}.
By changing the carrier density around $n=\pm n_{\mathtt{S}}/2$,
the resistance drops to zero, signaling the transition to a superconducting
state for temperatures below $T\approx1.0\,\text{K}$ \citep{Cao2018,Yankowitz2019}.
Conductance measurements also allowed to detect insulating states
at the fractional fillings $n/n_{\mathtt{S}}=1/4,3/4$ \citep{Yankowitz2019,Sharpe2019,Serlin2019,Polshyn2019},
and for a larger angle, $\theta\simeq1.27^{\circ}$, after applying
pressure \citep{Yankowitz2019}. Transport has further been essential
to inspect twisted double bilayers \citep{Cao2019,Shen2019}, to analyze
the role of Coulomb screening in both insulating and superconducting
phases of tBLG \citep{Liu2020}, and to demonstrate atomic reconstruction
for $\theta<1^{\circ}$ \citep{Yoo2019}. 

Despite its experimental relevance, there are comparatively few theoretical
works addressing transport in graphene heterostructures, in particular
on twisted bilayer systems. Ref.~\citep{Andelkovic2018} has addressed
the linear transport regime, using the Kubo formalism, finding that
a finite concentration of vacancies suppress the conductivity in a
wide energy region. Refs.~\citep{Hwang2019a,Wu2019c} discuss impurity
and phonon scattering, with particular emphasis on the temperature
dependence of the resistivity.Within the ballistic regime, an earlier
work addressed the transport properties of small nanoribbons with
a large angle tBLG section \citep{Pelc2015} highlighting the importance
of edge effects. Recently, a wider single layer graphene nanoribbon
with a twisted graphene flake on top was studied in Ref.~\citep{Bahamon2019}.
An interesting orbital magnetic structure has been predicted at finite
source-drain voltage applied only to the graphene nanoribbon. This
effect is attributed to the presence of counterflow currents first
discussed in \citep{bistritzerPNAS2011}. Finally, the effect of a
spatially inhomogeneous twist angle - twist disorder - on transport
has recently been considered in Ref. \citep{Padhi2020}. However,
the ballistic conductance of tBLG as a function of the twist angle
had , to our knowledge, not yet been obtained. Nonetheless, given
the high quality of the samples and the reported low temperatures,
the ballistic regime ought to be considered to make contact with the
experimental findings.

In this paper, we provide a thorough study of the ballistic transport
properties of tBLG as a function of the twist angle. We find three
qualitatively different regimes, characterizing the dependence of
the conductance on $\theta$, respectively for small, intermediate
and large angles. We pay particular attention to the behavior of conductance
near commensurate and incommensurate angles. In the large angle regime,
the results are shown to be sensitive to commensurability effects.
For intermediate angles, conductance features correlate with the position
of the van Hove singularities in the density of states (DOS). In the
small angle regime, commensurability effects completely disappear.
Superlattice induced gaps are clearly resolved, in agreement with
experimental findings.

The structure of the paper is as follows: In Sec.~\ref{sec:Model-and-methods}
we introduce the model and describe the methodology used to calculate
the conductance and the DOS. In Sec.~\ref{sec:Results-and-Discussion},
representative results of the transmission for the different angle
regimes are presented and discussed. Section~\ref{sec:Conclusion}
contains a short summary and the conclusions.

\section{Model and methods}

\label{sec:Model-and-methods}

\begin{figure}
\begin{centering}
\includegraphics[width=0.9\columnwidth]{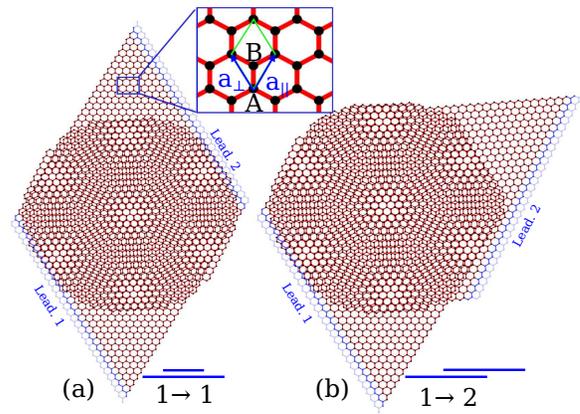}
\par\end{centering}
\centering{}\caption{\label{fig1: model}Twisted bilayer setups used in this work: (a)
$1\rightarrow1$ and (b) $1\rightarrow2$ setup. Vanishing blue color
denotes the beginning of semi-infinite leads. Lateral views are shown
under each setup. The typical linear size of the scattering region
is hundred of nanometers.}
\end{figure}

We consider transport in two setups, shown schematically in Fig.~\ref{fig1: model}.
The setup in (a) is denoted $1\rightarrow1$ and consists of a disk
of single layer graphene overlaid on a graphene ribbon. The diameter
of the disk is the same as the width of the ribbon, and its orientation
is chosen so that the two layers define a circularly shaped tBLG region
with the desired twist angle. The second setup, (b) denoted $1\rightarrow2$,
consists of two semi-infinite graphene ribbons with overlapping ends.
Semi-circular edges at the end of each ribbon define an almost circular
tBLG region. By adjusting the relative orientation of the ribbons
we fix the twist angle. In both setups, the remaining of the monolayer
ribbons define semi-infinite leads.

We model $p_{z}$ electrons in tBLG through a microscopic tight binding
Hamiltonian reading $H=H_{1}+H_{2}+H_{\perp}$, where $H_{l}$, with
$l=1,2$, is the Hamiltonian for a single layer, and $H_{\perp}$
is the interlayer coupling. For the single layer Hamiltonian we write,

\begin{equation}
H_{l}=-t\sum_{\underset{\vert\mathbf{R}_{l}-\mathbf{R}_{l}^{\prime}\vert\leq a}{\mathbf{R}_{l},\mathbf{R}_{l}^{\prime}}}c_{l,\alpha}^{\dagger}\left(\mathbf{R}_{l}\right)c_{l,\beta}\left(\mathbf{R}_{l}^{\prime}\right)\,,\label{eq:Hlayer}
\end{equation}
where $c_{l,\alpha}^{\dagger}\left(\mathbf{R}_{l}\right)$ creates
an electron at the Bravais lattice position $\mathbf{R}_{l}$, of
layer $l$ and sublattice $\alpha=A,B$, and the constraint $\vert\mathbf{R}_{l}-\mathbf{R}_{l}^{\prime}\vert\leq a$,
with $a=1.42\,\text{Å}$ for the carbon-carbon bond length, ensures
nearest neighbor hopping in each layer. The interlayer coupling is
written as $H_{\perp}=H_{12}+H_{21}$, with

\begin{equation}
H_{12}=\sum_{\mathbf{R}_{1},\mathbf{R}_{2}}t_{12}^{\alpha\beta}\left(\mathbf{R}_{1},\mathbf{R}_{2}\right)c_{1,\alpha}^{\dagger}\left(\mathbf{R}_{1}\right)c_{2,\beta}\left(\mathbf{R}_{2}\right)\,,\label{eq:Hperp}
\end{equation}
where $t_{12}^{\alpha\beta}\left(\mathbf{R}_{1},\mathbf{R}_{2}\right)$
is the interlayer hopping in the tight binding basis, and $H_{21}$
obtained from the previous equation by replacing $1\leftrightarrow2$.To
parametrize the interlayer hopping in Eq.~\eqref{eq:Hperp}, we use
the two-center approximation and assume $t_{12}^{\alpha\beta}\left(\mathbf{R}_{1},\mathbf{R}_{2}\right)$
depends only on the distance $r$ between the two $p_{z}$-orbitals,
\[
t_{12}^{\alpha\beta}\left(\mathbf{R}_{1},\mathbf{R}_{2}\right)\equiv t_{12}^{\alpha\beta}(r)\,,
\]
with $r^{2}=d_{\parallel}^{2}+d_{\perp}^{2}$ for an interlayer separation
$d_{\perp}=3.35\,\text{Å}$ and an in-plane projected distance $d_{\parallel}=\left|\mathbf{R}_{1}+\boldsymbol{\tau}_{1,\alpha}-\mathbf{R}_{2}-\boldsymbol{\tau}_{2,\beta}\right|$
between, where $\boldsymbol{\tau}_{l,\alpha}$ are the in-plane positions
of the orbital centers in the unit cell of each layer. Since the sites
in $A$ and $B$ sublattices correspond to the same $p_{z}$ orbital,
we further assume that the interlayer hopping does not depend on the
sublattice index, $t_{12}^{\alpha\beta}(r)\equiv t_{\perp}(r)$. Using
the Slater-Koster parametrization \citep{Slater1954}, we write

\begin{equation}
t_{\perp}\left(r\right)=\cos^{2}\left(\gamma\right)V_{pp\sigma}\left(r\right)+\sin^{2}\left(\gamma\right)V_{pp\pi}\left(r\right)\,,
\end{equation}
where the angle $\gamma$ is such that $\cos^{2}\left(\gamma\right)=d_{\perp}^{2}/r^{2}$,
and following Ref.~\citep{TramblyDeLaissardiere2012} the spatial
dependence of the parameters is given by

\begin{align}
V_{pp\sigma}\left(r\right)= & t_{\perp}\exp\left[q_{\sigma}\left(1-\frac{r}{d_{\perp}}\right)\right],\nonumber \\
V_{pp\pi}\left(r\right)= & -t\exp\left[q_{\pi}\left(1-\frac{r}{a}\right)\right].\label{eq: bond integrals}
\end{align}
From the second neighbor intralayer hopping, $t'=0.1t$, we fix $q_{\pi}=3.15$,
and assuming $q_{\pi}/a=q_{\sigma}/d_{\perp}$ yields $q_{\sigma}=7.42$
\citep{Goncalo2019}. For the remaining parameters, we consider $t=2.79\,\text{eV}$
and $t_{\perp}=0.35\,\text{eV}$.The sum in $H_{12}$ and $H_{21}$
is restricted to sites such that the in-plane seperation between sites
is smaller than a certain cutoff. We have considered only interlayer
hopping terms with $d_{\parallel}<0.9a$. Including further distance
interlayer hopping terms does not qualitatively alter the description
of tBLG \citep{LopesDosSantos2012}. One advantage of the truncation
we have chosen is that the first magic angle occurs at a slightly
larger value, $\theta^{*}\approx1.6^{\circ}$, thus for a slightly
smaller moiré \emph{cell}.

To compute the conductance, $G$, we use the Landauer approach. At
zero temperature and in the linear regime, the conductance is proportional
to the transmission, $T\left(\epsilon\right)$, for a system with
chemical potential $\epsilon$. We define $\bar{T}\left(\epsilon\right)=T\left(\epsilon\right)/w$,
where $w$ is the width of the leads, and write the conductance as
$G=G_{0}\bar{T}\left(\epsilon\right)$, where $G_{0}=\frac{2e^{2}}{h}$
is the conductance quantum. For both setups shown in Fig.~\ref{fig1: model},
the circular tBLG region with diameter $w$ defines the scattering
region, which, by construction, is connected to two semi-infinite
leads of width $w$. We compute the transmission, $T\left(\epsilon\right)$,
from the left lead to the right lead using the Kwant package \citep{Groth2014}.
Conductance calculations are performed with a scattering region containing
around $\mathcal{N}=3.77\times10^{5}$ carbon atoms for twist angles
$2^{\circ}<\theta<58^{\circ}$. For structures with smaller twist
angles, $0<\theta\leq2^{\circ}$ (close to AB stacking)  and $58\leq\theta<60^{\circ}$
(close to AA stacking), larger system sizes were used, with roughly
four times the numer of carbon atoms, i.e. around $\mathcal{N}=1.56\times10^{6}$.
The number of moiré \emph{cells} in the scattering region ranges from
$\gtrsim150$ for the smallest angles to several thousands for larger
angles. DOS calculations for incommensurate angles are done based
on the kernel polynomial method, also provided in Kwant. In this case,
the size of the tBLG circular region was kept at $\mathcal{N}=1.56\times10^{6}$
carbon atoms for angles in the range $2^{\circ}<\theta<58^{\circ}$,
and $\mathcal{N}=4.6\times10^{6}$ carbon atoms for smaller angles,
$0<\theta\leq2^{\circ}$ and $58\leq\theta<60^{\circ}.$ The DOS and
band structures for infinite tBLG systems was also evaluated using
a plane wave expansion for (incommensurate structures) and diagonalization
of the Bloch Hamiltonian (for commensurate structures).

\section{Results and Discussion}

\label{sec:Results-and-Discussion}

In this section, we present results for the conductance through a
tBLG region as a function of the twist angle for both setups shown
in Fig.~\ref{fig1: model}. We start with the analysis of general
features regarding the dependence on twist angle, and then move on
to discuss in greater detail the three distinct regimes: large, intermediate,
and small angles.

\begin{figure}
\begin{centering}
\includegraphics[width=0.95\columnwidth]{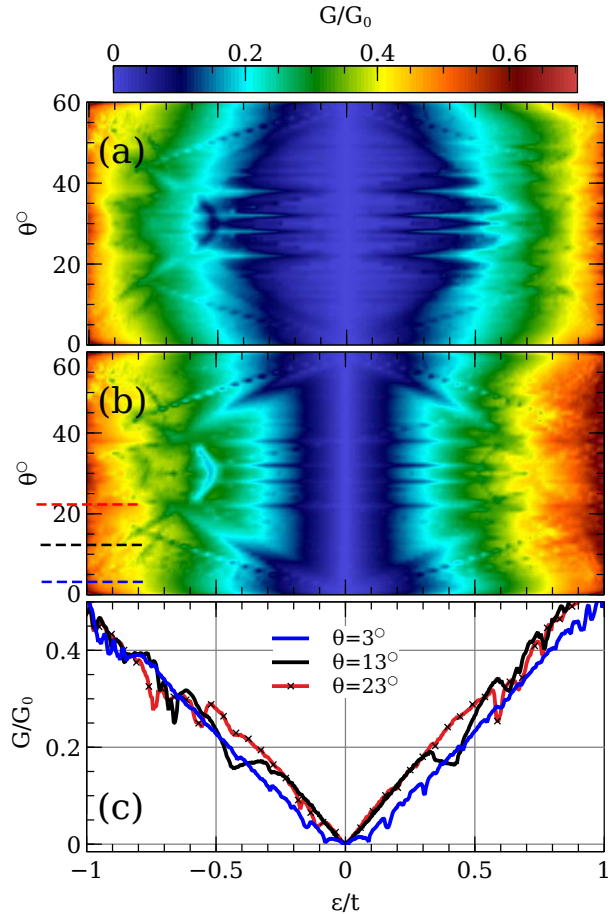}
\par\end{centering}
\centering{}\caption{\label{fig:Ggeneral}Density plot of the conductance, $G$, as a function
of twist angle, $\theta$, and energy, $\varepsilon$, for the two
setups considered in this work, $1\rightarrow2$~(a) and $1\rightarrow1$~(b).
Panel~(c) shows $G$ vs $\varepsilon$ for three particular angles,
$\theta=3^{\circ},\,13^{\circ}$ and $23^{\circ}$, depicted as dashed
horizontal lines in panel~(b). }
\end{figure}

\subsection{General features}

The conductance, $G$, of tBLG as a function of twist angle, $\theta$,
for an energy window $\varepsilon\in[-t,t]$, is shown in Figs.~\ref{fig:Ggeneral}(a)
and \ref{fig:Ggeneral}(b) for the two setups $1\rightarrow2$ and
$1\rightarrow1$, respectively. Generally, the conductance is lower
at lower energies, increasing non monotonically with energy for both
setups, irrespectively of the angle. For $\theta=0^{\circ}$, we recover
AB-stacked bilayer graphene where this general trend has previously
been observed \citep{Olyaei2019}. This behavior occurs even if the
scattering region is replaced by graphene itself, in which case it
is totally determined by the increasing number of channels in the
leads, i.e. proportional to the DOS.

In Fig.~\ref{fig:Ggeneral}(c) the conductance of the $1\rightarrow1$
setup is shown as a function of energy for three twist angles, representative
of the large, intermediate and small twists {[}corresponding to the
three horizontal dashed lines in Fig.~\ref{fig:Ggeneral}(b){]}.
Apart from the general trend of increasing conductance with increasing
energy, characteristic features of tBLG for small, intermediate, and
large angles can be appreciated. For small angles, the conductance
is suppressed within a wider energy region around $\varepsilon=0$
when compared with the other two angles. This low energy behavior
is associated with the flatband regime in tBLG, and will be further
discussed below. Intermediate and large twist angles have similar
$G(\varepsilon)$ at low energies. As will be shown below, at higher
energies the intermediate angle case first deviates from the large
angle behavior at the characteristic energy of the Von-Hove singularities
in tBLG. In the large angle regime, particular features of tBLG are
better appreciated as a function of twist angle. A series of peaks
and deeps can be seen in Figs.~\ref{fig:Ggeneral}(a) and \ref{fig:Ggeneral}(b).
As will be discussed below, these features appear at commensurate
twist angles with relatively small unit cells.

\subsection{Large twist angles}

\begin{figure}
\begin{centering}
\includegraphics[width=0.95\columnwidth]{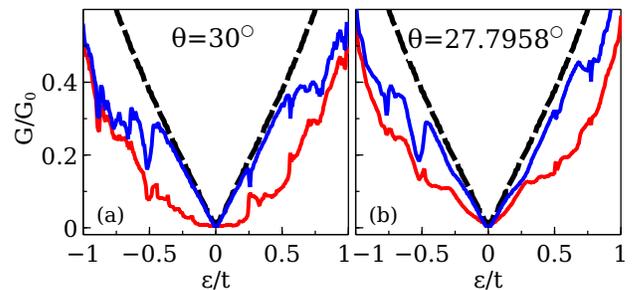}
\par\end{centering}
\centering{}\caption{\label{fig:GlargeAngle}Conductance, $G$, as a function of energy,
$\varepsilon$, for the $1\rightarrow1$ (blue) and $1\rightarrow2$
(red) setups computed at two large angles: an incommensurate structure
with twist angle $\theta=30^{\circ}$ (a) and a commensurate structure
with $\theta\simeq27.7958^{\circ}$ (b). The black dashed curve is
the conductance of graphene.}
\end{figure}

For large, incommensurate twist angles and low energies ($\varepsilon\ll t$),
the two layers become effectively decoupled: the conductance of the
$1\rightarrow1$ setup approaches that of monolayer graphene, while
for the $1\rightarrow2$ setup it is strongly suppressed as the electrons
need to tunnel to the other layer to conduct. This can be seen in
particular for $\theta=30^{\circ}$ in Fig.~\ref{fig:GlargeAngle}(a),
where we also show the graphene conductance as a dashed line for comparison.
The low energy decoupling of large angle tBLG has also been obtained
with the Kubo formalism within the linear response regime in Ref.~\citep{Andelkovic2018}.
However, at higher energies, deviations from the single layer conductance
for the $1\rightarrow1$ setup and an increasing conductance for the
$1\rightarrow2$ setup can be seen in Fig.~\ref{fig:GlargeAngle}(a).
This indicates that there is always some remnant coupling even for
the largest angles, in agreement with the observations of Ref.~\citep{SuarezMorell2011}.

The $\theta=30^{\circ}$ tBLG considered above is an example of an
incommensurate structure where no true Bravais lattice can be identified,
despite the presence of the moiré period. In particular, for the twist
angle $\theta=30^{\circ}$, tBLG has been identified as a new type
of quasicrystalline lattice with 12-fold rotational symmetry \citep{VanderLinden2012,Moon2019}.
For commensurate large angles, however, the conductance of tBLG behaves
differently. This can be seen by directly inspecting Fig.~\ref{fig:GlargeAngle}(b),
where we show the conductance $G(\varepsilon)$ for a commensurate
angle close to $\theta=30^{\circ}$. For the $1\rightarrow1$ setup,
$G(\varepsilon)$ starts to deviate from single layer graphene at
lower energies, and for the $1\rightarrow2$ setup, the low energy
conductance is not suppressed as for incommensurate angles.

\begin{figure}
\begin{centering}
\includegraphics[width=0.98\columnwidth]{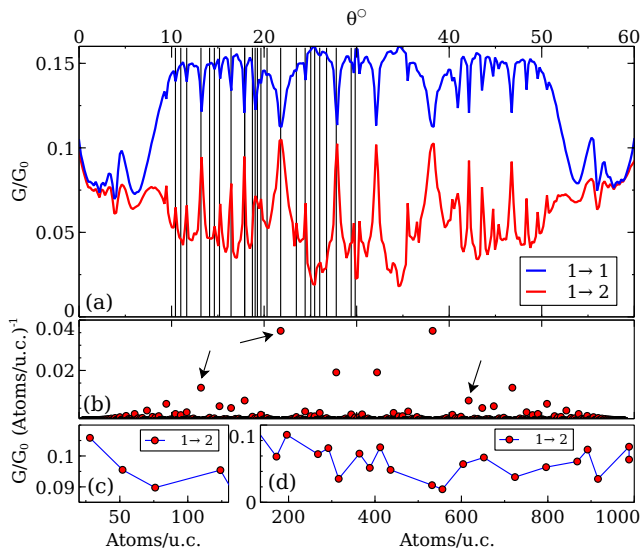}
\par\end{centering}
\centering{}\caption{\label{fig:GlargeCommens} Conductance, $G$, as a function of twist
angle, $\theta$, for the $1\rightarrow1$ (blue) and $1\rightarrow2$
(red) setups calculated at the energy $\varepsilon=0.225t$ (a). The
black vertical lines depict the position of commensurate angles, as
given by Eq.~\eqref{eq:commensCondition} in the main text. Panel
(b) exhibits the inverse of the size of the Wigner-Seitz unit cell
for commensurate structures versus the angle. Panels (c-d) show the
conductance values of the $1\rightarrow2$ setup for commensurate
structures as a function of size of the Wigner-Seitz unit cell for
small and large unit cells respectively. \textcolor{orange}{}}
\end{figure}

The difference between incommensurate and commensurate structures
at large twist angles is better appreciated in Fig.~\ref{fig:GlargeCommens}(a),
where we plot the conductance as a function of twist angle at the
fixed energy $\varepsilon=0.225t$ (easily reached via back gate field
effect). For the $1\rightarrow1$ setup a series of deeps and for
the $1\rightarrow2$ setup a series of peaks are clearly seen. These
are the same peaks and deeps observed in Figs.~\ref{fig:Ggeneral}(a)
and \ref{fig:Ggeneral}(b). As shown in Fig.~\ref{fig:GlargeCommens}(a),
the peaks/deeps match perfectly the vertical lines. It is also apparent
that the conductance curves are approximately symmetric with respect
to the twist angle $\theta=30^{\circ}$, and for that reason we only
plot vertical lines for $\theta<30^{\circ}$. The vertical lines correspond
to a series of commensurate angles, obtained according to the relation
\begin{equation}
\sin\left(\frac{\theta_{mr}}{2}\right)=\frac{r}{2\sqrt{3m^{2}+3mr+r^{2}}}\,,\label{eq:commensCondition}
\end{equation}
where $m\text{ and }r$ are positive coprime numbers \citep{LopesDosSantos2012}.
Therefore, the peaks/deeps are conductance signatures of commensurability.
They are robust to relative shifts of the layers (pure translations
with fixed $\theta$), and in Fig.~\ref{fig:GlargeCommens}(a) we
have averaged over shifts. A few of the largest peaks for the $1\rightarrow2$
setup have been reported also in the incoherent transport regime \citep{bistritzer2010transport}.
In the ballistic regime we see that the structure is very rich, with
peaks at many commensurate angles. In fact, we are lead to speculate
that peaks (deeps) may be present at every commensurate angle, though
its relative height (depth) may hinder the observation of most of
them. This is corroborated by an apparent correlation between the
height (depth) of the peak (deep) and the size of the Wigner-Seitz
unit cell for the corresponding commensurate lattice structure, as
shown in Fig.~\ref{fig:GlargeCommens}(b). There, the inverse of
the size of the Wigner-Seitz unit cell, measured in terms of the number
of atoms inside the cell, is plotted as a function of the respective
commensurate angle. By inspection it can be seen that the higher the
conductance peak in Fig.~\ref{fig:GlargeCommens}(a) for the $1\rightarrow2$
setup (the lower the deep for the $1\rightarrow1$ setup) the smaller
the respective Wigner-Seitz unit cell. The six peaks reported in Ref.~\citep{bistritzer2010transport}
for the incoherent regime correspond precisely to the six twist angles
with smallest Wigner-Seitz unit cell (higher conductance).

The conductance as a function of the size of the Wigner-Seitz unit
cell for commensurate structures is shown in Fig.~\ref{fig:GlargeCommens}(c).
In the large angle regime (unit cell sizes $\lesssim100$), it is
clear that the conductance decreases as the size of the cell increases.
For intermediate to low angles (unit cell sizes $\gtrsim100$), this
commensurability effect is lost. This agrees with the fact that $r=1$
structures, for which the unit cell coincides with the Moiré cell,
are special in the small-angle limit and determine the physics of
all types of commensurate structures \citep{LopesDosSantos2012}.
For very small angles, all commensurate structures are almost periodic
repetitions of structures with $r=1$.

\begin{figure}
\begin{centering}
\includegraphics[width=0.95\columnwidth]{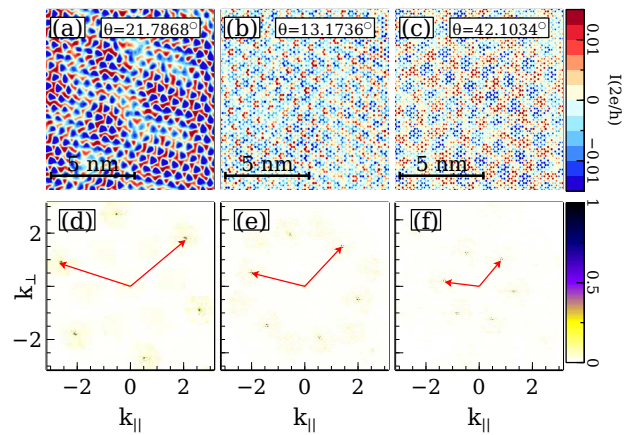}
\par\end{centering}
\centering{}\caption{\label{fig:current} (a-c) Interlayer current mapped onto the bottom
layer as a density plot at $\varepsilon=0.225t$ for three commensurate
structures. In terms of the integers $m$ and $r$ in Eq.~\eqref{eq:commensCondition},
$(m=1,r=1)$ in~(a) and in~(b), $(m=1,r=4)$ in~(c). (d-f) 2D
Fourier transform of the interlayer current shown in (a-c) as a function
of $k_{||}$ and $k_{\perp}$, respectively the longitudinal and transverse
momenta relative to the direction of the bottom lead ribbon. The arrows
are two representative vectors of the first star of reciprocal lattice
vectors for the respective commensurate structure.}
\end{figure}

In order to better understand the large angle commensurability effect,
we have computed the interlayer local current measured from the bottom
layer to the top layer in the $1\rightarrow2$ setup. To that purpose,
the bond current operator between a bottom layer site and a top layer
site was evaluated at the energy $\varepsilon=0.225t$. All the contributions
that connect to a given site in the bottom layer were then added up
and the obtained local interlayer current was assigned to that bottom
layer site. The corresponding map is shown in Figs.~\ref{fig:current}(a-c)
for three different commensurate angles {[}marked with an arrow in
Fig.~\ref{fig:GlargeCommens}(b){]}. Positive and negative values
mean interlayer current flowing in and out of the given bottom layer
site. The presence of a periodic pattern is apparent for the three
angles, as can be seen in Figs.~\ref{fig:current}(a-c). We took
the Fourier transform of the interlayer current by considering a rhombus
with $50\times50$ unit cells in the scattering region. The corresponding
map is shown in Figs.~\ref{fig:current}(d-f), respectively for the
three angles considered. The Fourier transform is extremely peaked,
as inferred from the very small dark dots on a whitish background
in Figs.~\ref{fig:current}(d-f). The peaks exactly fall onto the
first star of reciprocal lattice vectors for the respective commensurate
structure. Two representative vectors out of the six in the first
star are indicated in each panel of Fig.~\ref{fig:current}(d-f).
It follows that the observed periodicity for the interlayer current
in Figs.~\ref{fig:current}(a-c) mimics the periodicity of the commensurate
structure. The picture that emerges is that each Wigner-Seitz unit
cell contributes roughly the same to the interlayer current, so that
a higher conductance is obtained for a higher number of Wigner-Seitz
unit cells in he scattering region. Keeping the number of atoms in
the scattering region roughly the same, commensurate angles with smaller
Wigner-Seitz unit cells should have higher conductance, as observed.

\subsection{Intermediate twist angles}

\begin{figure}
\begin{centering}
\includegraphics[width=0.95\columnwidth]{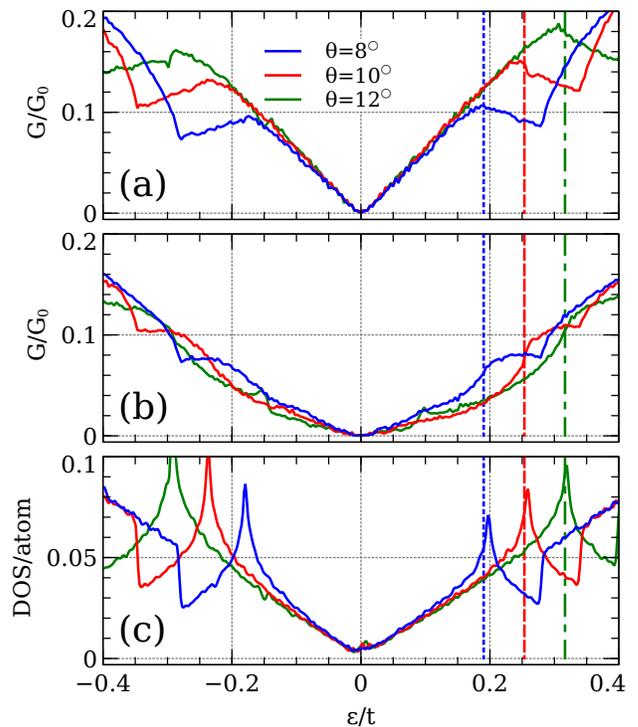}
\par\end{centering}
\centering{}\caption{\label{fig:GintermediateAngle}Conductance, $G$, as a function of
energy, $\varepsilon$, at three twist angles $\theta=8,10,12^{\circ}$,
for the $1\rightarrow1$~(a) and $1\rightarrow2$~(b) setups. The
corresponding DOS is shown in (c). Vertical dashed lines depict the
positions of Van Hove singularities computed through the approximate
analytical expression given in Eq.~\eqref{eq:VHSpos} for the positive
energy side.}
\end{figure}

\textcolor{black}{Figure~\ref{fig:GintermediateAngle}(a) and~\ref{fig:GintermediateAngle}(b)
shows the conductance at three representative angles in the intermediate
twist angle regime, respectively for the $1\rightarrow1$ and $1\rightarrow2$
setups and energies $|\varepsilon|<0.4t$. A salient feature in this
regime is the shoulder like behavior of the conductance around two
particular energies roughly symmetric around zero. The behavior is
more pronounced in the $1\rightarrow1$ setup {[}Fig.~\ref{fig:GintermediateAngle}(a){]}
which has a higher conductance, but it is clearly present in both
setups. The energy scale associated with this feature correlates with
the position of the two Van Hove singularities characteristic of tBLG
at moderate twist angles. This is clearly seen with the help of Fig.~\ref{fig:GintermediateAngle}(c),
where the DOS of the system (scattering region) is shown for the three
considered twist angles. The beginning of the shoulder-like feature
just signals the strong suppression of the DOS after the Van Hove
singularity. A similar effect is known to happen in single layer graphene
after the Van Hove singularity \citep{Olyaei2019}, though at much
higher energies.}

The presence of \textcolor{black}{Van Hove singularities in the DOS
of tBLG for the moderate twist angle regime originates from saddle
points in the energy dispersion. These saddle points are easily understood
as a consequence of the hybridization between single layer Dirac cones
}\citep{dSPN,LopesDosSantos2012}\textcolor{black}{. Due to rotation
by the twist angle $\theta$, and considering for the moment uncoupled
layers with $t_{\perp}=0$, the single layer Dirac cones appear separated
in reciprocal space by a distance }$\Delta K=2K\sin\left(\frac{\theta}{2}\right)$,
with $K=4\pi/\left(3\sqrt{3}a\right)$. Turning on the interlayer
coupling $t_{\perp}$, an avoided crossing at the energy scale $\varepsilon=\pm\hbar v_{f}\Delta K/2$
gives rise to saddle points at the approximate energies

\begin{equation}
\varepsilon_{vh}\approx\pm\left(\hbar v_{f}\frac{\Delta K}{2}-\frac{t_{\perp}}{2}\right).\label{eq:VHSpos}
\end{equation}
In \textcolor{black}{Fig.~\ref{fig:GintermediateAngle} the vertical
dashed lines are obtained through Eq.~}\eqref{eq:VHSpos} for the
three twist angles considered. The beginning of the shoulder-like
feature in the conductance as we increase energy is very well captured
by the energy scale $\varepsilon_{vh}$, as can be seen in \textcolor{black}{Figs.~\ref{fig:GintermediateAngle}}(a-b).

\subsection{Small twist angles}

It is in the small angle regime that most of the interesting novel
phases have been found \citep{Cao2018a,Cao2018,Chung2018,Yankowitz2019,Sharpe2019,Kerelsky2018,Choi2019,Jiang2019,Xie2019,Tomarken2019,Lu2019,Codecido2019,Shi2019,Serlin2019},
associated to the presence of extremely narrow bands at low energies.
Given the importance of transport measurements in accessing these
phases, and the fact that the model we use here is considered a proper
single particle description of tBLG, we address the question of what
are the conductance characteristics for this model at low twist angles.
The first result is that in the small angle regime $1\rightarrow1$
and $1\rightarrow2$ setups have very similar conductance at low energies.
This is an indication that in this regime the scattering region is
dominated by tBLG low energy properties, which weakens the differences
between the two setups. This is to be expected whenever the scattering
region is big enough to include a considerable number of moiré cells,
as is the case for the angles we consider. In the following we show
results only for the $1\rightarrow1$ setup.

\begin{figure}
\begin{centering}
\includegraphics[width=0.95\columnwidth]{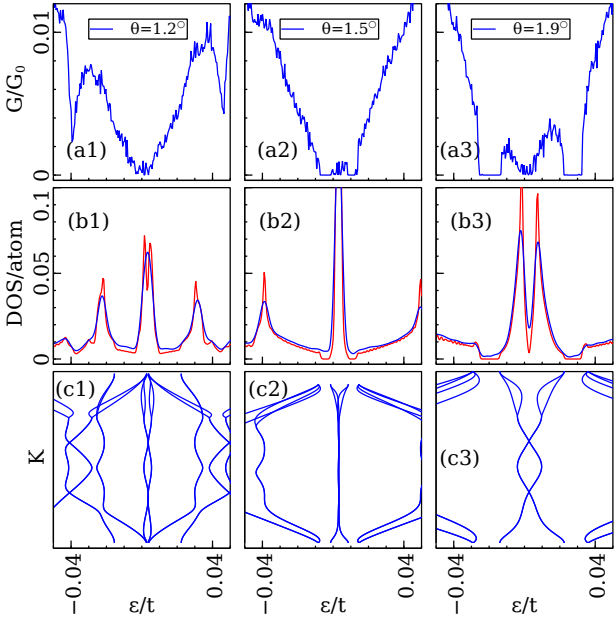}
\par\end{centering}
\centering{}\caption{\label{fig:GsmallAngleIncomm} (a1-a3) Conductance, $G$, as a function
of energy, $\varepsilon$, at three incommensurate twist angles around
the first \emph{magic angle} $\theta^{*}\approx1.6^{\circ}$ for the
$1\rightarrow1$ setup. (b1-b3) DOS of the scattering region used
in (a1-a3) is represented in blue. The DOS obtained using the plane
wave expansion method is shown in red. (c1-c3) Band structure obtained
by the plane wave expansion method.}
\end{figure}

We start with incommensurate angles. In Fig.~\ref{fig:GsmallAngleIncomm}(a1-a3)
the conductance at three representative small twist angles around
the flat band regime is shown for energies $|\varepsilon|\lesssim0.04t$.
For the model considered in this work, the flat band regime occurs
at $\theta^{*}\approx1.6^{\circ}$, so that in Fig.~\ref{fig:GsmallAngleIncomm}(a1)
the conductance if for an angle slightly below $\theta^{*}$, in Fig.~\ref{fig:GsmallAngleIncomm}(a2)
very close to $\theta^{*}$, and in \ref{fig:GsmallAngleIncomm}(a3)
slightly above. To confirm that the scattering region is indeed displaying
tBLG behavior around the flat band regime we have calculated the DOS
for the same angles, shown in Fig.~\ref{fig:GsmallAngleIncomm}(b1-b3)
as blue curves,using the Kwant package \citep{Groth2014}. The peaked
DOS around zero energy, particularly for $\theta\approx\theta^{*}$,
is a signature of tBLG behavior. Further confirmation comes from the
DOS obtained with the plane wave expansion method \citep{Goncalo2019},
suitable for incommensurate structures \citep{Amorim2018,Amorim2018b},
shown as red curves in Fig.~\ref{fig:GsmallAngleIncomm}(b1-b3).
. The agreement is remarkable despite the fact that Kwant is a construction
in real space using kernel polynomial methods while the plane wave
expansion method works in reciprocal space. The plane wave expansion
method allows also for the calculation of the band structure, which
is displayed in Fig.~\ref{fig:GsmallAngleIncomm}(c1-c3). There,
the appearance of very narrow bands at low energies, which become
especially flat near $\theta^{*}$, is apparent.

The conductance in Figs.~\ref{fig:GsmallAngleIncomm}(a2) and~\ref{fig:GsmallAngleIncomm}(a3)
has a low energy behavior which is not found at intermediate or large
twist angles, nor in single or AB-stacked bilayer graphene \citep{Olyaei2019}:
low energy finite conductance flanked by transport gaps where the
conductance vanishes. This is perfectly seen at $\theta^{*}$ and
when we approach $\theta^{*}$ from above. Comparison with DOS and
band structure, respectively in Figs.~\ref{fig:GsmallAngleIncomm}(b)
and~\ref{fig:GsmallAngleIncomm}(c), shows that the transport gaps
correlate perfectly with spectral gaps surrounding the low energy
narrow bands. These transport gaps are in perfect agreement with those
obtained experimentally when the low energy narrow bands are completely
occupied, for a carrier density $n=+n_{\mathtt{S}}$, or fully empty,
$n=-n_{\mathtt{S}}$ \citep{Cao2018a,Cao2018,Yankowitz2019,Sharpe2019,Serlin2019,Polshyn2019}.
Note also the similarity between the inverted double-well-like conductance
seen at low energies for $\theta\simeq2^{\circ}$ in Fig.~\ref{fig:GsmallAngleIncomm}(a3)
and the measurements of Ref.~\citep{jarilloHerrero2016} (compare
with conductivity) and Refs.~\citep{Kim2016,Chung2018} (compare
with inverse resistance). This points to the conclusion that the observed
behavior at small angles $\theta\gtrsim\theta^{*}$ is still captured
by the single particle description. Only very close to the \emph{magic
angle} $\theta^{*}$ are correlation effects expected to become relevant.
For angles $\theta\lesssim\theta^{*}$, as the one shown in Fig.~\ref{fig:GsmallAngleIncomm}(a1),
the conductance still correlates well with the electronic structure.
In particular, the two side peaks seen in the DOS of Fig.~\ref{fig:GsmallAngleIncomm}(b1)
clearly match the beginning of a shoulder like feature in conductance
as energy increases in absolute value. As can be appreciated in Fig.~\ref{fig:GsmallAngleIncomm}(c1),
the low energy narrow bands are no longer well isolated from the other
bands. This is the reason why there are no transport gaps in the conductance
at low energy.

\begin{figure}
\begin{centering}
\includegraphics[width=0.95\columnwidth]{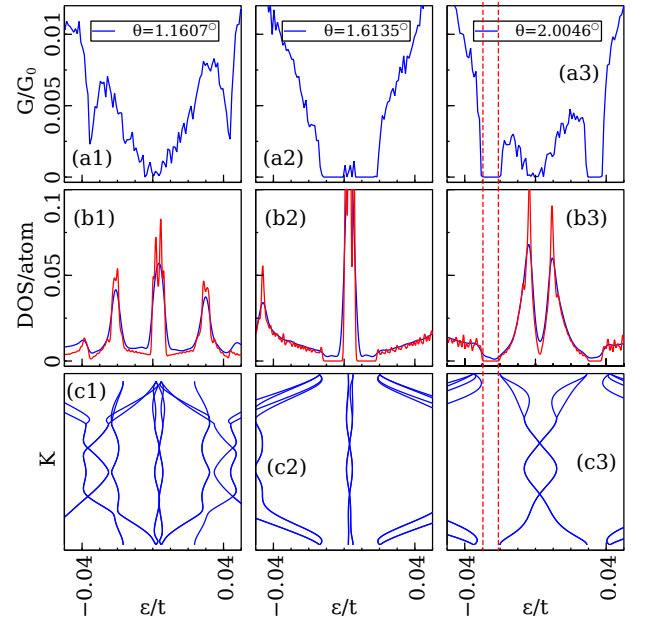}
\par\end{centering}
\centering{}\caption{\label{fig:GsmallAngleComm}(a1-a3) Conductance, $G$, as a function
of energy, $\varepsilon$, at three commensurate twist angles around
the first \emph{magic angle} $\theta^{*}\approx1.6^{\circ}$ for the
$1\rightarrow1$ setup. The three structures are $(m=1,r=1)$ in terms
of the integers $m$ and $r$ in Eq.~\eqref{eq:commensCondition}.
(b1-b3) DOS of the scattering region used in (a1-a3) is represented
in blue. The DOS obtained using the diagonalization of the Bloch Hamiltonian
is shown in red. (c1-c3) Band structure obtained by diagonalization
of the Bloch Hamiltonian.}
\end{figure}

The results for commensurate angles are shown in Fig.~\ref{fig:GsmallAngleComm}.
There is a close similarity with the results obtained for nearby commensurate
angles, presented in Fig.~\ref{fig:GsmallAngleIncomm}. Commensurability
effects, if present in the small twist angle regime, are significantly
milder than at large angles. Note in particular how the transport
gaps perfectly correlate with the spectrum, as is indicated by the
vertical lines on the right panels of Fig.~\ref{fig:GsmallAngleComm}.
We note \emph{en passant} that the DOS of the scattering region, shown
in blue in Fig.~\ref{fig:GsmallAngleComm}(b), does not vanish in
the energy region where transport gaps occur {[}this also happens
for the incommensurate angles shown in Fig.~\ref{fig:GsmallAngleIncomm}{]}.
However, by diagonalization of the Bloch Hamiltonian we obtain a vanishing
DOS (in red) compatible the the spectral gap shown in Fig.~\ref{fig:GsmallAngleComm}(c).
The reason is due to the open edges of the scattering region where
localized states, which do not contribute to conductance, exist.

\section{Conclusions}

\label{sec:Conclusion}

We have characterized ballistic charge transport across a twisted
bilayer graphene region connected to single-layer graphene leads.
We analyzed two setups: a flake of graphene on top of an infinite
graphene ribbon, dubbed $1\to1$, and two overlapping semi-infinite
graphene ribbons, dubbed $1\to2$. As a function of twist angle, we
found a strong dependence and identified three qualitatively different
regimes. For large angles, there are pronounced fluctuations and strong
commensurability effects: for generic incommensurate angles the two
graphene layers effectively decouple yielding regions of high (low)
conductance for the $1\to1$ ($1\to2$) geometry; large commensurate
angles corresponding to a small Wigner-Seitz unit cell appear as sharp
dips (peaks) in the conductance for the $1\to1$ ($1\to2$) case,
for which the two layers are strongly coupled. For intermediate angles,
we have found a correlation of the conductance features with the low
energy Van Hove singularities of the DOS. This suggests that conductance
measurements can be used as a measurement tool to determine the twist
angle. Finally, for small angles, commensurate effects seem to be
washed out. The almost flat bands appearing in this regime give rise
to distinctive conductance features that correlate with those found
in the DOS. In this regime, our results agree with the recent experimental
findings where transport gaps have been correlated with spectral gaps
\citep{jarilloHerrero2016,Kim2016,Cao2018a,Cao2018,Yankowitz2019,Sharpe2019,Serlin2019,Polshyn2019,Chung2018}.
\begin{acknowledgments}
HZO acknowledges the support from the DP-PMI and FCT-Portugal through
scholarship PD/BD/113649/2015. B.A. acknowledges funding from the
European Union\textquoteright s Horizon 2020 research and innovation
program under Grant Agreement No.~706538, and funding from FCT-Portugal
through Grant No.~CEECIND/02936/2017. EVC acknowledges support from
FCT-Portugal through Grant No.~UIDB/04650/2020. PR and HZO acknowledge
support by FCT-Portugal through Grant No.~UID/CTM/04540/2019. Some
computations were performed on the Baltasar-Sete-Sóis cluster, supported
by V. Cardoso\textquoteright s H2020 ERC Consolidator Grant no. MaGRaTh-646597,
computer assistance was provided by CENTRA/IST.
\end{acknowledgments}

\appendix
\bibliographystyle{apsrev4-1}
\bibliography{phd_2nd_paper}

\begin{thebibliography}{57}%
\makeatletter
\providecommand \@ifxundefined [1]{%
 \@ifx{#1\undefined}
}%
\providecommand \@ifnum [1]{%
 \ifnum #1\expandafter \@firstoftwo
 \else \expandafter \@secondoftwo
 \fi
}%
\providecommand \@ifx [1]{%
 \ifx #1\expandafter \@firstoftwo
 \else \expandafter \@secondoftwo
 \fi
}%
\providecommand \natexlab [1]{#1}%
\providecommand \enquote  [1]{``#1''}%
\providecommand \bibnamefont  [1]{#1}%
\providecommand \bibfnamefont [1]{#1}%
\providecommand \citenamefont [1]{#1}%
\providecommand \href@noop [0]{\@secondoftwo}%
\providecommand \href [0]{\begingroup \@sanitize@url \@href}%
\providecommand \@href[1]{\@@startlink{#1}\@@href}%
\providecommand \@@href[1]{\endgroup#1\@@endlink}%
\providecommand \@sanitize@url [0]{\catcode `\\12\catcode `\$12\catcode
  `\&12\catcode `\#12\catcode `\^12\catcode `\_12\catcode `\%12\relax}%
\providecommand \@@startlink[1]{}%
\providecommand \@@endlink[0]{}%
\providecommand \url  [0]{\begingroup\@sanitize@url \@url }%
\providecommand \@url [1]{\endgroup\@href {#1}{\urlprefix }}%
\providecommand \urlprefix  [0]{URL }%
\providecommand \Eprint [0]{\href }%
\providecommand \doibase [0]{http://dx.doi.org/}%
\providecommand \selectlanguage [0]{\@gobble}%
\providecommand \bibinfo  [0]{\@secondoftwo}%
\providecommand \bibfield  [0]{\@secondoftwo}%
\providecommand \translation [1]{[#1]}%
\providecommand \BibitemOpen [0]{}%
\providecommand \bibitemStop [0]{}%
\providecommand \bibitemNoStop [0]{.\EOS\space}%
\providecommand \EOS [0]{\spacefactor3000\relax}%
\providecommand \BibitemShut  [1]{\csname bibitem#1\endcsname}%
\let\auto@bib@innerbib\@empty
\bibitem [{\citenamefont {Cao}\ \emph {et~al.}(2018{\natexlab{a}})\citenamefont
  {Cao}, \citenamefont {Fatemi}, \citenamefont {Demir}, \citenamefont {Fang},
  \citenamefont {Tomarken}, \citenamefont {Luo}, \citenamefont
  {Sanchez-Yamagishi}, \citenamefont {Watanabe}, \citenamefont {Taniguchi},
  \citenamefont {Kaxiras}, \citenamefont {Ashoori},\ and\ \citenamefont
  {Jarillo-Herrero}}]{Cao2018a}%
  \BibitemOpen
  \bibfield  {author} {\bibinfo {author} {\bibfnamefont {Y.}~\bibnamefont
  {Cao}}, \bibinfo {author} {\bibfnamefont {V.}~\bibnamefont {Fatemi}},
  \bibinfo {author} {\bibfnamefont {A.}~\bibnamefont {Demir}}, \bibinfo
  {author} {\bibfnamefont {S.}~\bibnamefont {Fang}}, \bibinfo {author}
  {\bibfnamefont {S.~L.}\ \bibnamefont {Tomarken}}, \bibinfo {author}
  {\bibfnamefont {J.~Y.}\ \bibnamefont {Luo}}, \bibinfo {author} {\bibfnamefont
  {J.~D.}\ \bibnamefont {Sanchez-Yamagishi}}, \bibinfo {author} {\bibfnamefont
  {K.}~\bibnamefont {Watanabe}}, \bibinfo {author} {\bibfnamefont
  {T.}~\bibnamefont {Taniguchi}}, \bibinfo {author} {\bibfnamefont
  {E.}~\bibnamefont {Kaxiras}}, \bibinfo {author} {\bibfnamefont {R.~C.}\
  \bibnamefont {Ashoori}}, \ and\ \bibinfo {author} {\bibfnamefont
  {P.}~\bibnamefont {Jarillo-Herrero}},\ }\href {\doibase 10.1038/nature26154}
  {\bibfield  {journal} {\bibinfo  {journal} {Nature}\ }\textbf {\bibinfo
  {volume} {556}},\ \bibinfo {pages} {80} (\bibinfo {year}
  {2018}{\natexlab{a}})},\ \Eprint {http://arxiv.org/abs/1802.00553}
  {arXiv:1802.00553} \BibitemShut {NoStop}%
\bibitem [{\citenamefont {Cao}\ \emph {et~al.}(2018{\natexlab{b}})\citenamefont
  {Cao}, \citenamefont {Fatemi}, \citenamefont {Fang}, \citenamefont
  {Watanabe}, \citenamefont {Taniguchi}, \citenamefont {Kaxiras},\ and\
  \citenamefont {Jarillo-Herrero}}]{Cao2018}%
  \BibitemOpen
  \bibfield  {author} {\bibinfo {author} {\bibfnamefont {Y.}~\bibnamefont
  {Cao}}, \bibinfo {author} {\bibfnamefont {V.}~\bibnamefont {Fatemi}},
  \bibinfo {author} {\bibfnamefont {S.}~\bibnamefont {Fang}}, \bibinfo {author}
  {\bibfnamefont {K.}~\bibnamefont {Watanabe}}, \bibinfo {author}
  {\bibfnamefont {T.}~\bibnamefont {Taniguchi}}, \bibinfo {author}
  {\bibfnamefont {E.}~\bibnamefont {Kaxiras}}, \ and\ \bibinfo {author}
  {\bibfnamefont {P.}~\bibnamefont {Jarillo-Herrero}},\ }\href {\doibase
  10.1038/nature26160} {\bibfield  {journal} {\bibinfo  {journal} {Nature}\
  }\textbf {\bibinfo {volume} {556}},\ \bibinfo {pages} {43} (\bibinfo {year}
  {2018}{\natexlab{b}})},\ \Eprint {http://arxiv.org/abs/1803.02342}
  {arXiv:1803.02342} \BibitemShut {NoStop}%
\bibitem [{\citenamefont {{Lopes dos Santos}}\ \emph
  {et~al.}(2007)\citenamefont {{Lopes dos Santos}}, \citenamefont {Peres},\
  and\ \citenamefont {{Castro Neto}}}]{dSPN}%
  \BibitemOpen
  \bibfield  {author} {\bibinfo {author} {\bibfnamefont {J.~M.~B.}\
  \bibnamefont {{Lopes dos Santos}}}, \bibinfo {author} {\bibfnamefont
  {N.~M.~R.}\ \bibnamefont {Peres}}, \ and\ \bibinfo {author} {\bibfnamefont
  {A.~H.}\ \bibnamefont {{Castro Neto}}},\ }\href {\doibase
  10.1103/PhysRevLett.99.256802} {\bibfield  {journal} {\bibinfo  {journal}
  {Phys. Rev. Lett.}\ }\textbf {\bibinfo {volume} {99}},\ \bibinfo {pages}
  {256802} (\bibinfo {year} {2007})},\ \Eprint {http://arxiv.org/abs/0704.2128}
  {arXiv:0704.2128} \BibitemShut {NoStop}%
\bibitem [{\citenamefont {Bistritzer}\ and\ \citenamefont
  {MacDonald}(2011)}]{bistritzerPNAS2011}%
  \BibitemOpen
  \bibfield  {author} {\bibinfo {author} {\bibfnamefont {R.}~\bibnamefont
  {Bistritzer}}\ and\ \bibinfo {author} {\bibfnamefont {A.~H.}\ \bibnamefont
  {MacDonald}},\ }\href {\doibase 10.1073/pnas.1108174108} {\bibfield
  {journal} {\bibinfo  {journal} {Proc. Natl. Acad. Sci.}\ }\textbf {\bibinfo
  {volume} {108}},\ \bibinfo {pages} {12233} (\bibinfo {year}
  {2011})}\BibitemShut {NoStop}%
\bibitem [{\citenamefont {{Lopes dos Santos}}\ \emph
  {et~al.}(2012)\citenamefont {{Lopes dos Santos}}, \citenamefont {Peres},\
  and\ \citenamefont {{Castro Neto}}}]{LopesDosSantos2012}%
  \BibitemOpen
  \bibfield  {author} {\bibinfo {author} {\bibfnamefont {J.~M.~B.}\
  \bibnamefont {{Lopes dos Santos}}}, \bibinfo {author} {\bibfnamefont
  {N.~M.~R.}\ \bibnamefont {Peres}}, \ and\ \bibinfo {author} {\bibfnamefont
  {A.~H.}\ \bibnamefont {{Castro Neto}}},\ }\href {\doibase
  10.1103/PhysRevB.86.155449} {\bibfield  {journal} {\bibinfo  {journal} {Phys.
  Rev. B}\ }\textbf {\bibinfo {volume} {86}},\ \bibinfo {pages} {155449}
  (\bibinfo {year} {2012})},\ \Eprint {http://arxiv.org/abs/1202.1088}
  {arXiv:1202.1088} \BibitemShut {NoStop}%
\bibitem [{\citenamefont {{De Trambly Laissardi{\`{e}}re}}\ \emph
  {et~al.}(2010)\citenamefont {{De Trambly Laissardi{\`{e}}re}}, \citenamefont
  {Mayou},\ and\ \citenamefont {Magaud}}]{TramblydeLaissardiere2010a}%
  \BibitemOpen
  \bibfield  {author} {\bibinfo {author} {\bibfnamefont {G.}~\bibnamefont {{De
  Trambly Laissardi{\`{e}}re}}}, \bibinfo {author} {\bibfnamefont
  {D.}~\bibnamefont {Mayou}}, \ and\ \bibinfo {author} {\bibfnamefont
  {L.}~\bibnamefont {Magaud}},\ }\href {\doibase 10.1021/nl902948m} {\bibfield
  {journal} {\bibinfo  {journal} {Nano Lett.}\ }\textbf {\bibinfo {volume}
  {10}},\ \bibinfo {pages} {804} (\bibinfo {year} {2010})},\ \Eprint
  {http://arxiv.org/abs/0904.1233} {arXiv:0904.1233} \BibitemShut {NoStop}%
\bibitem [{\citenamefont {{Su{\'{a}}rez Morell}}\ \emph
  {et~al.}(2010)\citenamefont {{Su{\'{a}}rez Morell}}, \citenamefont {Correa},
  \citenamefont {Vargas}, \citenamefont {Pacheco},\ and\ \citenamefont
  {Barticevic}}]{SuarezMorell2010}%
  \BibitemOpen
  \bibfield  {author} {\bibinfo {author} {\bibfnamefont {E.}~\bibnamefont
  {{Su{\'{a}}rez Morell}}}, \bibinfo {author} {\bibfnamefont {J.~D.}\
  \bibnamefont {Correa}}, \bibinfo {author} {\bibfnamefont {P.}~\bibnamefont
  {Vargas}}, \bibinfo {author} {\bibfnamefont {M.}~\bibnamefont {Pacheco}}, \
  and\ \bibinfo {author} {\bibfnamefont {Z.}~\bibnamefont {Barticevic}},\
  }\href {\doibase 10.1103/PhysRevB.82.121407} {\bibfield  {journal} {\bibinfo
  {journal} {Phys. Rev. B}\ }\textbf {\bibinfo {volume} {82}},\ \bibinfo
  {pages} {121407} (\bibinfo {year} {2010})},\ \Eprint
  {http://arxiv.org/abs/1012.4320} {arXiv:1012.4320} \BibitemShut {NoStop}%
\bibitem [{\citenamefont {Shallcross}\ \emph {et~al.}(2010)\citenamefont
  {Shallcross}, \citenamefont {Sharma}, \citenamefont {Kandelaki},\ and\
  \citenamefont {Pankratov}}]{Shallcross2010}%
  \BibitemOpen
  \bibfield  {author} {\bibinfo {author} {\bibfnamefont {S.}~\bibnamefont
  {Shallcross}}, \bibinfo {author} {\bibfnamefont {S.}~\bibnamefont {Sharma}},
  \bibinfo {author} {\bibfnamefont {E.}~\bibnamefont {Kandelaki}}, \ and\
  \bibinfo {author} {\bibfnamefont {O.~A.}\ \bibnamefont {Pankratov}},\ }\href
  {\doibase 10.1103/PhysRevB.81.165105} {\bibfield  {journal} {\bibinfo
  {journal} {Phys. Rev. B - Condens. Matter Mater. Phys.}\ }\textbf {\bibinfo
  {volume} {81}},\ \bibinfo {pages} {165105} (\bibinfo {year}
  {2010})}\BibitemShut {NoStop}%
\bibitem [{\citenamefont {Chung}\ \emph {et~al.}(2018)\citenamefont {Chung},
  \citenamefont {Xu},\ and\ \citenamefont {Chen}}]{Chung2018}%
  \BibitemOpen
  \bibfield  {author} {\bibinfo {author} {\bibfnamefont {T.-F.}\ \bibnamefont
  {Chung}}, \bibinfo {author} {\bibfnamefont {Y.}~\bibnamefont {Xu}}, \ and\
  \bibinfo {author} {\bibfnamefont {Y.~P.}\ \bibnamefont {Chen}},\ }\href
  {\doibase 10.1103/PhysRevB.98.035425} {\bibfield  {journal} {\bibinfo
  {journal} {Phys. Rev. B}\ }\textbf {\bibinfo {volume} {98}},\ \bibinfo
  {pages} {035425} (\bibinfo {year} {2018})}\BibitemShut {NoStop}%
\bibitem [{\citenamefont {Yankowitz}\ \emph {et~al.}(2019)\citenamefont
  {Yankowitz}, \citenamefont {Chen}, \citenamefont {Polshyn}, \citenamefont
  {Zhang}, \citenamefont {Watanabe}, \citenamefont {Taniguchi}, \citenamefont
  {Graf}, \citenamefont {Young},\ and\ \citenamefont {Dean}}]{Yankowitz2019}%
  \BibitemOpen
  \bibfield  {author} {\bibinfo {author} {\bibfnamefont {M.}~\bibnamefont
  {Yankowitz}}, \bibinfo {author} {\bibfnamefont {S.}~\bibnamefont {Chen}},
  \bibinfo {author} {\bibfnamefont {H.}~\bibnamefont {Polshyn}}, \bibinfo
  {author} {\bibfnamefont {Y.}~\bibnamefont {Zhang}}, \bibinfo {author}
  {\bibfnamefont {K.}~\bibnamefont {Watanabe}}, \bibinfo {author}
  {\bibfnamefont {T.}~\bibnamefont {Taniguchi}}, \bibinfo {author}
  {\bibfnamefont {D.}~\bibnamefont {Graf}}, \bibinfo {author} {\bibfnamefont
  {A.~F.}\ \bibnamefont {Young}}, \ and\ \bibinfo {author} {\bibfnamefont
  {C.~R.}\ \bibnamefont {Dean}},\ }\href {\doibase 10.1126/science.aav1910}
  {\bibfield  {journal} {\bibinfo  {journal} {Science}\ }\textbf {\bibinfo
  {volume} {363}},\ \bibinfo {pages} {1059} (\bibinfo {year}
  {2019})}\BibitemShut {NoStop}%
\bibitem [{\citenamefont {Sharpe}\ \emph {et~al.}(2019)\citenamefont {Sharpe},
  \citenamefont {Fox}, \citenamefont {Barnard}, \citenamefont {Finney},
  \citenamefont {Watanabe}, \citenamefont {Taniguchi}, \citenamefont
  {Kastner},\ and\ \citenamefont {Goldhaber-Gordon}}]{Sharpe2019}%
  \BibitemOpen
  \bibfield  {author} {\bibinfo {author} {\bibfnamefont {A.~L.}\ \bibnamefont
  {Sharpe}}, \bibinfo {author} {\bibfnamefont {E.~J.}\ \bibnamefont {Fox}},
  \bibinfo {author} {\bibfnamefont {A.~W.}\ \bibnamefont {Barnard}}, \bibinfo
  {author} {\bibfnamefont {J.}~\bibnamefont {Finney}}, \bibinfo {author}
  {\bibfnamefont {K.}~\bibnamefont {Watanabe}}, \bibinfo {author}
  {\bibfnamefont {T.}~\bibnamefont {Taniguchi}}, \bibinfo {author}
  {\bibfnamefont {M.~A.}\ \bibnamefont {Kastner}}, \ and\ \bibinfo {author}
  {\bibfnamefont {D.}~\bibnamefont {Goldhaber-Gordon}},\ }\href {\doibase
  10.1126/science.aaw3780} {\bibfield  {journal} {\bibinfo  {journal}
  {Science}\ }\textbf {\bibinfo {volume} {365}},\ \bibinfo {pages} {605}
  (\bibinfo {year} {2019})},\ \Eprint {http://arxiv.org/abs/1901.03520}
  {arXiv:1901.03520} \BibitemShut {NoStop}%
\bibitem [{\citenamefont {Kerelsky}\ \emph {et~al.}(2019)\citenamefont
  {Kerelsky}, \citenamefont {McGilly}, \citenamefont {Kennes}, \citenamefont
  {Xian}, \citenamefont {Yankowitz}, \citenamefont {Chen}, \citenamefont
  {Watanabe}, \citenamefont {Taniguchi}, \citenamefont {Hone}, \citenamefont
  {Dean}, \citenamefont {Rubio},\ and\ \citenamefont
  {Pasupathy}}]{Kerelsky2018}%
  \BibitemOpen
  \bibfield  {author} {\bibinfo {author} {\bibfnamefont {A.}~\bibnamefont
  {Kerelsky}}, \bibinfo {author} {\bibfnamefont {L.~J.}\ \bibnamefont
  {McGilly}}, \bibinfo {author} {\bibfnamefont {D.~M.}\ \bibnamefont {Kennes}},
  \bibinfo {author} {\bibfnamefont {L.}~\bibnamefont {Xian}}, \bibinfo {author}
  {\bibfnamefont {M.}~\bibnamefont {Yankowitz}}, \bibinfo {author}
  {\bibfnamefont {S.}~\bibnamefont {Chen}}, \bibinfo {author} {\bibfnamefont
  {K.}~\bibnamefont {Watanabe}}, \bibinfo {author} {\bibfnamefont
  {T.}~\bibnamefont {Taniguchi}}, \bibinfo {author} {\bibfnamefont
  {J.}~\bibnamefont {Hone}}, \bibinfo {author} {\bibfnamefont {C.}~\bibnamefont
  {Dean}}, \bibinfo {author} {\bibfnamefont {A.}~\bibnamefont {Rubio}}, \ and\
  \bibinfo {author} {\bibfnamefont {A.~N.}\ \bibnamefont {Pasupathy}},\ }\href
  {\doibase 10.1038/s41586-019-1431-9} {\bibfield  {journal} {\bibinfo
  {journal} {Nature}\ }\textbf {\bibinfo {volume} {572}},\ \bibinfo {pages}
  {95} (\bibinfo {year} {2019})},\ \Eprint {http://arxiv.org/abs/1812.08776}
  {arXiv:1812.08776} \BibitemShut {NoStop}%
\bibitem [{\citenamefont {Choi}\ \emph {et~al.}(2019)\citenamefont {Choi},
  \citenamefont {Kemmer}, \citenamefont {Peng}, \citenamefont {Thomson},
  \citenamefont {Arora}, \citenamefont {Polski}, \citenamefont {Zhang},
  \citenamefont {Ren}, \citenamefont {Alicea}, \citenamefont {Refael},
  \citenamefont {von Oppen}, \citenamefont {Watanabe}, \citenamefont
  {Taniguchi},\ and\ \citenamefont {Nadj-Perge}}]{Choi2019}%
  \BibitemOpen
  \bibfield  {author} {\bibinfo {author} {\bibfnamefont {Y.}~\bibnamefont
  {Choi}}, \bibinfo {author} {\bibfnamefont {J.}~\bibnamefont {Kemmer}},
  \bibinfo {author} {\bibfnamefont {Y.}~\bibnamefont {Peng}}, \bibinfo {author}
  {\bibfnamefont {A.}~\bibnamefont {Thomson}}, \bibinfo {author} {\bibfnamefont
  {H.}~\bibnamefont {Arora}}, \bibinfo {author} {\bibfnamefont
  {R.}~\bibnamefont {Polski}}, \bibinfo {author} {\bibfnamefont
  {Y.}~\bibnamefont {Zhang}}, \bibinfo {author} {\bibfnamefont
  {H.}~\bibnamefont {Ren}}, \bibinfo {author} {\bibfnamefont {J.}~\bibnamefont
  {Alicea}}, \bibinfo {author} {\bibfnamefont {G.}~\bibnamefont {Refael}},
  \bibinfo {author} {\bibfnamefont {F.}~\bibnamefont {von Oppen}}, \bibinfo
  {author} {\bibfnamefont {K.}~\bibnamefont {Watanabe}}, \bibinfo {author}
  {\bibfnamefont {T.}~\bibnamefont {Taniguchi}}, \ and\ \bibinfo {author}
  {\bibfnamefont {S.}~\bibnamefont {Nadj-Perge}},\ }\href {\doibase
  10.1038/s41567-019-0606-5} {\bibfield  {journal} {\bibinfo  {journal} {Nat.
  Phys.}\ }\textbf {\bibinfo {volume} {15}},\ \bibinfo {pages} {1174} (\bibinfo
  {year} {2019})},\ \Eprint {http://arxiv.org/abs/1901.02997}
  {arXiv:1901.02997} \BibitemShut {NoStop}%
\bibitem [{\citenamefont {Jiang}\ \emph {et~al.}(2019)\citenamefont {Jiang},
  \citenamefont {Lai}, \citenamefont {Watanabe}, \citenamefont {Taniguchi},
  \citenamefont {Haule}, \citenamefont {Mao},\ and\ \citenamefont
  {Andrei}}]{Jiang2019}%
  \BibitemOpen
  \bibfield  {author} {\bibinfo {author} {\bibfnamefont {Y.}~\bibnamefont
  {Jiang}}, \bibinfo {author} {\bibfnamefont {X.}~\bibnamefont {Lai}}, \bibinfo
  {author} {\bibfnamefont {K.}~\bibnamefont {Watanabe}}, \bibinfo {author}
  {\bibfnamefont {T.}~\bibnamefont {Taniguchi}}, \bibinfo {author}
  {\bibfnamefont {K.}~\bibnamefont {Haule}}, \bibinfo {author} {\bibfnamefont
  {J.}~\bibnamefont {Mao}}, \ and\ \bibinfo {author} {\bibfnamefont {E.~Y.}\
  \bibnamefont {Andrei}},\ }\href {\doibase 10.1038/s41586-019-1460-4}
  {\bibfield  {journal} {\bibinfo  {journal} {Nature}\ }\textbf {\bibinfo
  {volume} {573}},\ \bibinfo {pages} {91} (\bibinfo {year} {2019})},\ \Eprint
  {http://arxiv.org/abs/1904.10153} {arXiv:1904.10153} \BibitemShut {NoStop}%
\bibitem [{\citenamefont {Xie}\ \emph {et~al.}(2019)\citenamefont {Xie},
  \citenamefont {Lian}, \citenamefont {J{\"{a}}ck}, \citenamefont {Liu},
  \citenamefont {Chiu}, \citenamefont {Watanabe}, \citenamefont {Taniguchi},
  \citenamefont {Bernevig},\ and\ \citenamefont {Yazdani}}]{Xie2019}%
  \BibitemOpen
  \bibfield  {author} {\bibinfo {author} {\bibfnamefont {Y.}~\bibnamefont
  {Xie}}, \bibinfo {author} {\bibfnamefont {B.}~\bibnamefont {Lian}}, \bibinfo
  {author} {\bibfnamefont {B.}~\bibnamefont {J{\"{a}}ck}}, \bibinfo {author}
  {\bibfnamefont {X.}~\bibnamefont {Liu}}, \bibinfo {author} {\bibfnamefont
  {C.-L.}\ \bibnamefont {Chiu}}, \bibinfo {author} {\bibfnamefont
  {K.}~\bibnamefont {Watanabe}}, \bibinfo {author} {\bibfnamefont
  {T.}~\bibnamefont {Taniguchi}}, \bibinfo {author} {\bibfnamefont {B.~A.}\
  \bibnamefont {Bernevig}}, \ and\ \bibinfo {author} {\bibfnamefont
  {A.}~\bibnamefont {Yazdani}},\ }\href {\doibase 10.1038/s41586-019-1422-x}
  {\bibfield  {journal} {\bibinfo  {journal} {Nature}\ }\textbf {\bibinfo
  {volume} {572}},\ \bibinfo {pages} {101} (\bibinfo {year} {2019})},\ \Eprint
  {http://arxiv.org/abs/1906.09274} {arXiv:1906.09274} \BibitemShut {NoStop}%
\bibitem [{\citenamefont {Tomarken}\ \emph {et~al.}(2019)\citenamefont
  {Tomarken}, \citenamefont {Cao}, \citenamefont {Demir}, \citenamefont
  {Watanabe}, \citenamefont {Taniguchi}, \citenamefont {Jarillo-Herrero},\ and\
  \citenamefont {Ashoori}}]{Tomarken2019}%
  \BibitemOpen
  \bibfield  {author} {\bibinfo {author} {\bibfnamefont {S.~L.}\ \bibnamefont
  {Tomarken}}, \bibinfo {author} {\bibfnamefont {Y.}~\bibnamefont {Cao}},
  \bibinfo {author} {\bibfnamefont {A.}~\bibnamefont {Demir}}, \bibinfo
  {author} {\bibfnamefont {K.}~\bibnamefont {Watanabe}}, \bibinfo {author}
  {\bibfnamefont {T.}~\bibnamefont {Taniguchi}}, \bibinfo {author}
  {\bibfnamefont {P.}~\bibnamefont {Jarillo-Herrero}}, \ and\ \bibinfo {author}
  {\bibfnamefont {R.~C.}\ \bibnamefont {Ashoori}},\ }\href {\doibase
  10.1103/PhysRevLett.123.046601} {\bibfield  {journal} {\bibinfo  {journal}
  {Phys. Rev. Lett.}\ }\textbf {\bibinfo {volume} {123}},\ \bibinfo {pages}
  {046601} (\bibinfo {year} {2019})},\ \Eprint
  {http://arxiv.org/abs/1903.10492} {arXiv:1903.10492} \BibitemShut {NoStop}%
\bibitem [{\citenamefont {Lu}\ \emph {et~al.}(2019)\citenamefont {Lu},
  \citenamefont {Stepanov}, \citenamefont {Yang}, \citenamefont {Xie},
  \citenamefont {Aamir}, \citenamefont {Das}, \citenamefont {Urgell},
  \citenamefont {Watanabe}, \citenamefont {Taniguchi}, \citenamefont {Zhang},
  \citenamefont {Bachtold}, \citenamefont {MacDonald},\ and\ \citenamefont
  {Efetov}}]{Lu2019}%
  \BibitemOpen
  \bibfield  {author} {\bibinfo {author} {\bibfnamefont {X.}~\bibnamefont
  {Lu}}, \bibinfo {author} {\bibfnamefont {P.}~\bibnamefont {Stepanov}},
  \bibinfo {author} {\bibfnamefont {W.}~\bibnamefont {Yang}}, \bibinfo {author}
  {\bibfnamefont {M.}~\bibnamefont {Xie}}, \bibinfo {author} {\bibfnamefont
  {M.~A.}\ \bibnamefont {Aamir}}, \bibinfo {author} {\bibfnamefont
  {I.}~\bibnamefont {Das}}, \bibinfo {author} {\bibfnamefont {C.}~\bibnamefont
  {Urgell}}, \bibinfo {author} {\bibfnamefont {K.}~\bibnamefont {Watanabe}},
  \bibinfo {author} {\bibfnamefont {T.}~\bibnamefont {Taniguchi}}, \bibinfo
  {author} {\bibfnamefont {G.}~\bibnamefont {Zhang}}, \bibinfo {author}
  {\bibfnamefont {A.}~\bibnamefont {Bachtold}}, \bibinfo {author}
  {\bibfnamefont {A.~H.}\ \bibnamefont {MacDonald}}, \ and\ \bibinfo {author}
  {\bibfnamefont {D.~K.}\ \bibnamefont {Efetov}},\ }\href {\doibase
  10.1038/s41586-019-1695-0} {\bibfield  {journal} {\bibinfo  {journal}
  {Nature}\ }\textbf {\bibinfo {volume} {574}},\ \bibinfo {pages} {653}
  (\bibinfo {year} {2019})},\ \Eprint {http://arxiv.org/abs/1903.06513}
  {arXiv:1903.06513} \BibitemShut {NoStop}%
\bibitem [{\citenamefont {Codecido}\ \emph {et~al.}(2019)\citenamefont
  {Codecido}, \citenamefont {Wang}, \citenamefont {Koester}, \citenamefont
  {Che}, \citenamefont {Tian}, \citenamefont {Lv}, \citenamefont {Tran},
  \citenamefont {Watanabe}, \citenamefont {Taniguchi}, \citenamefont {Zhang},
  \citenamefont {Bockrath},\ and\ \citenamefont {Lau}}]{Codecido2019}%
  \BibitemOpen
  \bibfield  {author} {\bibinfo {author} {\bibfnamefont {E.}~\bibnamefont
  {Codecido}}, \bibinfo {author} {\bibfnamefont {Q.}~\bibnamefont {Wang}},
  \bibinfo {author} {\bibfnamefont {R.}~\bibnamefont {Koester}}, \bibinfo
  {author} {\bibfnamefont {S.}~\bibnamefont {Che}}, \bibinfo {author}
  {\bibfnamefont {H.}~\bibnamefont {Tian}}, \bibinfo {author} {\bibfnamefont
  {R.}~\bibnamefont {Lv}}, \bibinfo {author} {\bibfnamefont {S.}~\bibnamefont
  {Tran}}, \bibinfo {author} {\bibfnamefont {K.}~\bibnamefont {Watanabe}},
  \bibinfo {author} {\bibfnamefont {T.}~\bibnamefont {Taniguchi}}, \bibinfo
  {author} {\bibfnamefont {F.}~\bibnamefont {Zhang}}, \bibinfo {author}
  {\bibfnamefont {M.}~\bibnamefont {Bockrath}}, \ and\ \bibinfo {author}
  {\bibfnamefont {C.~N.}\ \bibnamefont {Lau}},\ }\href {\doibase
  10.1126/sciadv.aaw9770} {\bibfield  {journal} {\bibinfo  {journal} {Sci.
  Adv.}\ }\textbf {\bibinfo {volume} {5}} (\bibinfo {year} {2019}),\
  10.1126/sciadv.aaw9770},\ \Eprint {http://arxiv.org/abs/1902.05151}
  {arXiv:1902.05151} \BibitemShut {NoStop}%
\bibitem [{\citenamefont {Shi}\ \emph {et~al.}(2020)\citenamefont {Shi},
  \citenamefont {Zhan}, \citenamefont {Qi}, \citenamefont {Huang},
  \citenamefont {van Veen}, \citenamefont {Silva-Guill{\'{e}}n}, \citenamefont
  {Zhang}, \citenamefont {Li}, \citenamefont {Xie}, \citenamefont {Ji},
  \citenamefont {Katsnelson}, \citenamefont {Yuan}, \citenamefont {Qin},\ and\
  \citenamefont {Zhang}}]{Shi2019}%
  \BibitemOpen
  \bibfield  {author} {\bibinfo {author} {\bibfnamefont {H.}~\bibnamefont
  {Shi}}, \bibinfo {author} {\bibfnamefont {Z.}~\bibnamefont {Zhan}}, \bibinfo
  {author} {\bibfnamefont {Z.}~\bibnamefont {Qi}}, \bibinfo {author}
  {\bibfnamefont {K.}~\bibnamefont {Huang}}, \bibinfo {author} {\bibfnamefont
  {E.}~\bibnamefont {van Veen}}, \bibinfo {author} {\bibfnamefont
  {J.~{\'{A}}.}\ \bibnamefont {Silva-Guill{\'{e}}n}}, \bibinfo {author}
  {\bibfnamefont {R.}~\bibnamefont {Zhang}}, \bibinfo {author} {\bibfnamefont
  {P.}~\bibnamefont {Li}}, \bibinfo {author} {\bibfnamefont {K.}~\bibnamefont
  {Xie}}, \bibinfo {author} {\bibfnamefont {H.}~\bibnamefont {Ji}}, \bibinfo
  {author} {\bibfnamefont {M.~I.}\ \bibnamefont {Katsnelson}}, \bibinfo
  {author} {\bibfnamefont {S.}~\bibnamefont {Yuan}}, \bibinfo {author}
  {\bibfnamefont {S.}~\bibnamefont {Qin}}, \ and\ \bibinfo {author}
  {\bibfnamefont {Z.}~\bibnamefont {Zhang}},\ }\href {\doibase
  10.1038/s41467-019-14207-w} {\bibfield  {journal} {\bibinfo  {journal} {Nat.
  Commun.}\ }\textbf {\bibinfo {volume} {11}},\ \bibinfo {pages} {371}
  (\bibinfo {year} {2020})},\ \Eprint {http://arxiv.org/abs/1905.04515}
  {arXiv:1905.04515} \BibitemShut {NoStop}%
\bibitem [{\citenamefont {Serlin}\ \emph {et~al.}(2020)\citenamefont {Serlin},
  \citenamefont {Tschirhart}, \citenamefont {Polshyn}, \citenamefont {Zhang},
  \citenamefont {Zhu}, \citenamefont {Watanabe}, \citenamefont {Taniguchi},
  \citenamefont {Balents},\ and\ \citenamefont {Young}}]{Serlin2019}%
  \BibitemOpen
  \bibfield  {author} {\bibinfo {author} {\bibfnamefont {M.}~\bibnamefont
  {Serlin}}, \bibinfo {author} {\bibfnamefont {C.~L.}\ \bibnamefont
  {Tschirhart}}, \bibinfo {author} {\bibfnamefont {H.}~\bibnamefont {Polshyn}},
  \bibinfo {author} {\bibfnamefont {Y.}~\bibnamefont {Zhang}}, \bibinfo
  {author} {\bibfnamefont {J.}~\bibnamefont {Zhu}}, \bibinfo {author}
  {\bibfnamefont {K.}~\bibnamefont {Watanabe}}, \bibinfo {author}
  {\bibfnamefont {T.}~\bibnamefont {Taniguchi}}, \bibinfo {author}
  {\bibfnamefont {L.}~\bibnamefont {Balents}}, \ and\ \bibinfo {author}
  {\bibfnamefont {A.~F.}\ \bibnamefont {Young}},\ }\href {\doibase
  10.1126/science.aay5533} {\bibfield  {journal} {\bibinfo  {journal}
  {Science}\ }\textbf {\bibinfo {volume} {367}},\ \bibinfo {pages} {900}
  (\bibinfo {year} {2020})},\ \Eprint {http://arxiv.org/abs/1907.00261}
  {arXiv:1907.00261} \BibitemShut {NoStop}%
\bibitem [{\citenamefont {Shen}\ \emph {et~al.}(2019)\citenamefont {Shen},
  \citenamefont {Li}, \citenamefont {Wang}, \citenamefont {Zhao}, \citenamefont
  {Tang}, \citenamefont {Liu}, \citenamefont {Tian}, \citenamefont {Chu},
  \citenamefont {Watanabe}, \citenamefont {Taniguchi}, \citenamefont {Yang},
  \citenamefont {Meng}, \citenamefont {Shi},\ and\ \citenamefont
  {Zhang}}]{Shen2019}%
  \BibitemOpen
  \bibfield  {author} {\bibinfo {author} {\bibfnamefont {C.}~\bibnamefont
  {Shen}}, \bibinfo {author} {\bibfnamefont {N.}~\bibnamefont {Li}}, \bibinfo
  {author} {\bibfnamefont {S.}~\bibnamefont {Wang}}, \bibinfo {author}
  {\bibfnamefont {Y.}~\bibnamefont {Zhao}}, \bibinfo {author} {\bibfnamefont
  {J.}~\bibnamefont {Tang}}, \bibinfo {author} {\bibfnamefont {J.}~\bibnamefont
  {Liu}}, \bibinfo {author} {\bibfnamefont {J.}~\bibnamefont {Tian}}, \bibinfo
  {author} {\bibfnamefont {Y.}~\bibnamefont {Chu}}, \bibinfo {author}
  {\bibfnamefont {K.}~\bibnamefont {Watanabe}}, \bibinfo {author}
  {\bibfnamefont {T.}~\bibnamefont {Taniguchi}}, \bibinfo {author}
  {\bibfnamefont {R.}~\bibnamefont {Yang}}, \bibinfo {author} {\bibfnamefont
  {Z.~Y.}\ \bibnamefont {Meng}}, \bibinfo {author} {\bibfnamefont
  {D.}~\bibnamefont {Shi}}, \ and\ \bibinfo {author} {\bibfnamefont
  {G.}~\bibnamefont {Zhang}},\ }\href {http://arxiv.org/abs/1903.06952} {\
  (\bibinfo {year} {2019})},\ \Eprint {http://arxiv.org/abs/1903.06952}
  {arXiv:1903.06952} \BibitemShut {NoStop}%
\bibitem [{\citenamefont {Liu}\ \emph {et~al.}(2019)\citenamefont {Liu},
  \citenamefont {Hao}, \citenamefont {Khalaf}, \citenamefont {Lee},
  \citenamefont {Watanabe}, \citenamefont {Taniguchi}, \citenamefont
  {Vishwanath},\ and\ \citenamefont {Kim}}]{Liu2019a}%
  \BibitemOpen
  \bibfield  {author} {\bibinfo {author} {\bibfnamefont {X.}~\bibnamefont
  {Liu}}, \bibinfo {author} {\bibfnamefont {Z.}~\bibnamefont {Hao}}, \bibinfo
  {author} {\bibfnamefont {E.}~\bibnamefont {Khalaf}}, \bibinfo {author}
  {\bibfnamefont {J.~Y.}\ \bibnamefont {Lee}}, \bibinfo {author} {\bibfnamefont
  {K.}~\bibnamefont {Watanabe}}, \bibinfo {author} {\bibfnamefont
  {T.}~\bibnamefont {Taniguchi}}, \bibinfo {author} {\bibfnamefont
  {A.}~\bibnamefont {Vishwanath}}, \ and\ \bibinfo {author} {\bibfnamefont
  {P.}~\bibnamefont {Kim}},\ }\href {https://arxiv.org/abs/1903.08130
  http://arxiv.org/abs/1903.08130} {\  (\bibinfo {year} {2019})},\ \Eprint
  {http://arxiv.org/abs/1903.08130} {arXiv:1903.08130} \BibitemShut {NoStop}%
\bibitem [{\citenamefont {Cao}\ \emph {et~al.}(2019)\citenamefont {Cao},
  \citenamefont {Rodan-Legrain}, \citenamefont {Rubies-Bigorda}, \citenamefont
  {Park}, \citenamefont {Watanabe}, \citenamefont {Taniguchi},\ and\
  \citenamefont {Jarillo-Herrero}}]{Cao2019}%
  \BibitemOpen
  \bibfield  {author} {\bibinfo {author} {\bibfnamefont {Y.}~\bibnamefont
  {Cao}}, \bibinfo {author} {\bibfnamefont {D.}~\bibnamefont {Rodan-Legrain}},
  \bibinfo {author} {\bibfnamefont {O.}~\bibnamefont {Rubies-Bigorda}},
  \bibinfo {author} {\bibfnamefont {J.~M.}\ \bibnamefont {Park}}, \bibinfo
  {author} {\bibfnamefont {K.}~\bibnamefont {Watanabe}}, \bibinfo {author}
  {\bibfnamefont {T.}~\bibnamefont {Taniguchi}}, \ and\ \bibinfo {author}
  {\bibfnamefont {P.}~\bibnamefont {Jarillo-Herrero}},\ }\href
  {http://arxiv.org/abs/1903.08596} {\  (\bibinfo {year} {2019})},\ \Eprint
  {http://arxiv.org/abs/1903.08596} {arXiv:1903.08596} \BibitemShut {NoStop}%
\bibitem [{\citenamefont {Chen}\ \emph {et~al.}(2019)\citenamefont {Chen},
  \citenamefont {Sharpe}, \citenamefont {Gallagher}, \citenamefont {Rosen},
  \citenamefont {Fox}, \citenamefont {Jiang}, \citenamefont {Lyu},
  \citenamefont {Li}, \citenamefont {Watanabe}, \citenamefont {Taniguchi},
  \citenamefont {Jung}, \citenamefont {Shi}, \citenamefont {Goldhaber-Gordon},
  \citenamefont {Zhang},\ and\ \citenamefont {Wang}}]{Chen2019}%
  \BibitemOpen
  \bibfield  {author} {\bibinfo {author} {\bibfnamefont {G.}~\bibnamefont
  {Chen}}, \bibinfo {author} {\bibfnamefont {A.~L.}\ \bibnamefont {Sharpe}},
  \bibinfo {author} {\bibfnamefont {P.}~\bibnamefont {Gallagher}}, \bibinfo
  {author} {\bibfnamefont {I.~T.}\ \bibnamefont {Rosen}}, \bibinfo {author}
  {\bibfnamefont {E.~J.}\ \bibnamefont {Fox}}, \bibinfo {author} {\bibfnamefont
  {L.}~\bibnamefont {Jiang}}, \bibinfo {author} {\bibfnamefont
  {B.}~\bibnamefont {Lyu}}, \bibinfo {author} {\bibfnamefont {H.}~\bibnamefont
  {Li}}, \bibinfo {author} {\bibfnamefont {K.}~\bibnamefont {Watanabe}},
  \bibinfo {author} {\bibfnamefont {T.}~\bibnamefont {Taniguchi}}, \bibinfo
  {author} {\bibfnamefont {J.}~\bibnamefont {Jung}}, \bibinfo {author}
  {\bibfnamefont {Z.}~\bibnamefont {Shi}}, \bibinfo {author} {\bibfnamefont
  {D.}~\bibnamefont {Goldhaber-Gordon}}, \bibinfo {author} {\bibfnamefont
  {Y.}~\bibnamefont {Zhang}}, \ and\ \bibinfo {author} {\bibfnamefont
  {F.}~\bibnamefont {Wang}},\ }\href {\doibase 10.1038/s41586-019-1393-y}
  {\bibfield  {journal} {\bibinfo  {journal} {Nature}\ }\textbf {\bibinfo
  {volume} {572}},\ \bibinfo {pages} {215} (\bibinfo {year}
  {2019})}\BibitemShut {NoStop}%
\bibitem [{\citenamefont {Zuo}\ \emph {et~al.}(2018)\citenamefont {Zuo},
  \citenamefont {Qiao}, \citenamefont {Ma}, \citenamefont {Yin}, \citenamefont
  {Sun}, \citenamefont {Zhang}, \citenamefont {Guan},\ and\ \citenamefont
  {He}}]{Zuo2018}%
  \BibitemOpen
  \bibfield  {author} {\bibinfo {author} {\bibfnamefont {W.-J.}\ \bibnamefont
  {Zuo}}, \bibinfo {author} {\bibfnamefont {J.-B.}\ \bibnamefont {Qiao}},
  \bibinfo {author} {\bibfnamefont {D.-L.}\ \bibnamefont {Ma}}, \bibinfo
  {author} {\bibfnamefont {L.-J.}\ \bibnamefont {Yin}}, \bibinfo {author}
  {\bibfnamefont {G.}~\bibnamefont {Sun}}, \bibinfo {author} {\bibfnamefont
  {J.-Y.}\ \bibnamefont {Zhang}}, \bibinfo {author} {\bibfnamefont {L.-Y.}\
  \bibnamefont {Guan}}, \ and\ \bibinfo {author} {\bibfnamefont
  {L.}~\bibnamefont {He}},\ }\href {\doibase 10.1103/PhysRevB.97.035440}
  {\bibfield  {journal} {\bibinfo  {journal} {Phys. Rev. B}\ }\textbf {\bibinfo
  {volume} {97}},\ \bibinfo {pages} {035440} (\bibinfo {year} {2018})},\
  \Eprint {http://arxiv.org/abs/1711.08109} {arXiv:1711.08109} \BibitemShut
  {NoStop}%
\bibitem [{\citenamefont {Wu}\ \emph {et~al.}(2019{\natexlab{a}})\citenamefont
  {Wu}, \citenamefont {Lovorn}, \citenamefont {Tutuc}, \citenamefont {Martin},\
  and\ \citenamefont {MacDonald}}]{Wu2019}%
  \BibitemOpen
  \bibfield  {author} {\bibinfo {author} {\bibfnamefont {F.}~\bibnamefont
  {Wu}}, \bibinfo {author} {\bibfnamefont {T.}~\bibnamefont {Lovorn}}, \bibinfo
  {author} {\bibfnamefont {E.}~\bibnamefont {Tutuc}}, \bibinfo {author}
  {\bibfnamefont {I.}~\bibnamefont {Martin}}, \ and\ \bibinfo {author}
  {\bibfnamefont {A.~H.}\ \bibnamefont {MacDonald}},\ }\href {\doibase
  10.1103/PhysRevLett.122.086402} {\bibfield  {journal} {\bibinfo  {journal}
  {Phys. Rev. Lett.}\ }\textbf {\bibinfo {volume} {122}},\ \bibinfo {pages}
  {086402} (\bibinfo {year} {2019}{\natexlab{a}})}\BibitemShut {NoStop}%
\bibitem [{\citenamefont {Wang}\ \emph {et~al.}(2019)\citenamefont {Wang},
  \citenamefont {Shih}, \citenamefont {Ghiotto}, \citenamefont {Xian},
  \citenamefont {Rhodes}, \citenamefont {Tan}, \citenamefont {Claassen},
  \citenamefont {Kennes}, \citenamefont {Bai}, \citenamefont {Kim},
  \citenamefont {Watanabe}, \citenamefont {Taniguchi}, \citenamefont {Zhu},
  \citenamefont {Hone}, \citenamefont {Rubio}, \citenamefont {Pasupathy},\ and\
  \citenamefont {Dean}}]{Wang2019}%
  \BibitemOpen
  \bibfield  {author} {\bibinfo {author} {\bibfnamefont {L.}~\bibnamefont
  {Wang}}, \bibinfo {author} {\bibfnamefont {E.-M.}\ \bibnamefont {Shih}},
  \bibinfo {author} {\bibfnamefont {A.}~\bibnamefont {Ghiotto}}, \bibinfo
  {author} {\bibfnamefont {L.}~\bibnamefont {Xian}}, \bibinfo {author}
  {\bibfnamefont {D.~A.}\ \bibnamefont {Rhodes}}, \bibinfo {author}
  {\bibfnamefont {C.}~\bibnamefont {Tan}}, \bibinfo {author} {\bibfnamefont
  {M.}~\bibnamefont {Claassen}}, \bibinfo {author} {\bibfnamefont {D.~M.}\
  \bibnamefont {Kennes}}, \bibinfo {author} {\bibfnamefont {Y.}~\bibnamefont
  {Bai}}, \bibinfo {author} {\bibfnamefont {B.}~\bibnamefont {Kim}}, \bibinfo
  {author} {\bibfnamefont {K.}~\bibnamefont {Watanabe}}, \bibinfo {author}
  {\bibfnamefont {T.}~\bibnamefont {Taniguchi}}, \bibinfo {author}
  {\bibfnamefont {X.}~\bibnamefont {Zhu}}, \bibinfo {author} {\bibfnamefont
  {J.}~\bibnamefont {Hone}}, \bibinfo {author} {\bibfnamefont {A.}~\bibnamefont
  {Rubio}}, \bibinfo {author} {\bibfnamefont {A.}~\bibnamefont {Pasupathy}}, \
  and\ \bibinfo {author} {\bibfnamefont {C.~R.}\ \bibnamefont {Dean}},\
  }\href@noop {} {\  (\bibinfo {year} {2019})},\ \Eprint
  {http://arxiv.org/abs/1910.12147} {arXiv:1910.12147} \BibitemShut {NoStop}%
\bibitem [{\citenamefont {Gon{\c{c}}alo}\ \emph {et~al.}(2019)\citenamefont
  {Gon{\c{c}}alo}, \citenamefont {Amorim}, \citenamefont {Castro},
  \citenamefont {Lopes},\ and\ \citenamefont {Peres}}]{Goncalo2019}%
  \BibitemOpen
  \bibfield  {author} {\bibinfo {author} {\bibfnamefont {C.}~\bibnamefont
  {Gon{\c{c}}alo}}, \bibinfo {author} {\bibfnamefont {B.}~\bibnamefont
  {Amorim}}, \bibinfo {author} {\bibfnamefont {E.~V.}\ \bibnamefont {Castro}},
  \bibinfo {author} {\bibfnamefont {J.~M. V.~P.}\ \bibnamefont {Lopes}}, \ and\
  \bibinfo {author} {\bibfnamefont {N.}~\bibnamefont {Peres}},\ }in\ \href@noop
  {} {\emph {\bibinfo {booktitle} {Handb. Graphene Vol. 3}}},\ \bibinfo
  {editor} {edited by\ \bibinfo {editor} {\bibfnamefont {M.}~\bibnamefont
  {Zhang}}}\ (\bibinfo  {publisher} {John Wiley {\&} Sons, New Jersey},\
  \bibinfo {year} {2019})\ Chap.~\bibinfo {chapter} {6}, pp.\ \bibinfo {pages}
  {177--230}\BibitemShut {NoStop}%
\bibitem [{\citenamefont {Yuan}\ \emph {et~al.}(2018)\citenamefont {Yuan},
  \citenamefont {Koshino}, \citenamefont {Fu}, \citenamefont {Koretsune},
  \citenamefont {Ochi},\ and\ \citenamefont {Kuroki}}]{Yuan2018a}%
  \BibitemOpen
  \bibfield  {author} {\bibinfo {author} {\bibfnamefont {N.~F.~Q.}\
  \bibnamefont {Yuan}}, \bibinfo {author} {\bibfnamefont {M.}~\bibnamefont
  {Koshino}}, \bibinfo {author} {\bibfnamefont {L.}~\bibnamefont {Fu}},
  \bibinfo {author} {\bibfnamefont {T.}~\bibnamefont {Koretsune}}, \bibinfo
  {author} {\bibfnamefont {M.}~\bibnamefont {Ochi}}, \ and\ \bibinfo {author}
  {\bibfnamefont {K.}~\bibnamefont {Kuroki}},\ }\href {\doibase
  10.1103/physrevx.8.031087} {\bibfield  {journal} {\bibinfo  {journal} {Phys.
  Rev. X}\ }\textbf {\bibinfo {volume} {8}},\ \bibinfo {pages} {031087}
  (\bibinfo {year} {2018})}\BibitemShut {NoStop}%
\bibitem [{\citenamefont {Zou}\ \emph {et~al.}(2018)\citenamefont {Zou},
  \citenamefont {Po}, \citenamefont {Vishwanath},\ and\ \citenamefont
  {Senthil}}]{Zou2018}%
  \BibitemOpen
  \bibfield  {author} {\bibinfo {author} {\bibfnamefont {L.}~\bibnamefont
  {Zou}}, \bibinfo {author} {\bibfnamefont {H.~C.}\ \bibnamefont {Po}},
  \bibinfo {author} {\bibfnamefont {A.}~\bibnamefont {Vishwanath}}, \ and\
  \bibinfo {author} {\bibfnamefont {T.}~\bibnamefont {Senthil}},\ }\href
  {\doibase 10.1103/PhysRevB.98.085435} {\bibfield  {journal} {\bibinfo
  {journal} {Phys. Rev. B}\ }\textbf {\bibinfo {volume} {98}},\ \bibinfo
  {pages} {085435} (\bibinfo {year} {2018})},\ \Eprint
  {http://arxiv.org/abs/1806.07873} {arXiv:1806.07873} \BibitemShut {NoStop}%
\bibitem [{\citenamefont {Kang}\ and\ \citenamefont {Vafek}(2018)}]{Kang2018}%
  \BibitemOpen
  \bibfield  {author} {\bibinfo {author} {\bibfnamefont {J.}~\bibnamefont
  {Kang}}\ and\ \bibinfo {author} {\bibfnamefont {O.}~\bibnamefont {Vafek}},\
  }\href {\doibase 10.1103/PhysRevX.8.031088} {\bibfield  {journal} {\bibinfo
  {journal} {Phys. Rev. X}\ }\textbf {\bibinfo {volume} {8}},\ \bibinfo {pages}
  {031088} (\bibinfo {year} {2018})}\BibitemShut {NoStop}%
\bibitem [{\citenamefont {Po}\ \emph {et~al.}(2018)\citenamefont {Po},
  \citenamefont {Zou}, \citenamefont {Senthil},\ and\ \citenamefont
  {Vishwanath}}]{Po2018a}%
  \BibitemOpen
  \bibfield  {author} {\bibinfo {author} {\bibfnamefont {H.~C.}\ \bibnamefont
  {Po}}, \bibinfo {author} {\bibfnamefont {L.}~\bibnamefont {Zou}}, \bibinfo
  {author} {\bibfnamefont {T.}~\bibnamefont {Senthil}}, \ and\ \bibinfo
  {author} {\bibfnamefont {A.}~\bibnamefont {Vishwanath}},\ }\href {\doibase
  10.1103/PhysRevB.99.195455} {\bibfield  {journal} {\bibinfo  {journal} {Phys.
  Rev. B}\ }\textbf {\bibinfo {volume} {99}},\ \bibinfo {pages} {195455}
  (\bibinfo {year} {2018})},\ \Eprint {http://arxiv.org/abs/1808.02482}
  {arXiv:1808.02482} \BibitemShut {NoStop}%
\bibitem [{\citenamefont {Yuan}\ and\ \citenamefont {Fu}(2018)}]{Yuan2018}%
  \BibitemOpen
  \bibfield  {author} {\bibinfo {author} {\bibfnamefont {N.~F.~Q.}\
  \bibnamefont {Yuan}}\ and\ \bibinfo {author} {\bibfnamefont {L.}~\bibnamefont
  {Fu}},\ }\href {\doibase 10.1103/PhysRevB.98.045103} {\bibfield  {journal}
  {\bibinfo  {journal} {Phys. Rev. B}\ }\textbf {\bibinfo {volume} {98}},\
  \bibinfo {pages} {045103} (\bibinfo {year} {2018})},\ \Eprint
  {http://arxiv.org/abs/1803.09699} {arXiv:1803.09699} \BibitemShut {NoStop}%
\bibitem [{\citenamefont {Carr}\ \emph {et~al.}(2019)\citenamefont {Carr},
  \citenamefont {Fang}, \citenamefont {Po}, \citenamefont {Vishwanath},\ and\
  \citenamefont {Kaxiras}}]{Carr2019a}%
  \BibitemOpen
  \bibfield  {author} {\bibinfo {author} {\bibfnamefont {S.}~\bibnamefont
  {Carr}}, \bibinfo {author} {\bibfnamefont {S.}~\bibnamefont {Fang}}, \bibinfo
  {author} {\bibfnamefont {H.~C.}\ \bibnamefont {Po}}, \bibinfo {author}
  {\bibfnamefont {A.}~\bibnamefont {Vishwanath}}, \ and\ \bibinfo {author}
  {\bibfnamefont {E.}~\bibnamefont {Kaxiras}},\ }\href {\doibase
  10.1103/PhysRevResearch.1.033072} {\bibfield  {journal} {\bibinfo  {journal}
  {Phys. Rev. Res.}\ }\textbf {\bibinfo {volume} {1}},\ \bibinfo {pages}
  {033072} (\bibinfo {year} {2019})},\ \Eprint
  {http://arxiv.org/abs/1907.06282} {arXiv:1907.06282} \BibitemShut {NoStop}%
\bibitem [{\citenamefont {Shallcross}\ \emph {et~al.}(2013)\citenamefont
  {Shallcross}, \citenamefont {Sharma},\ and\ \citenamefont
  {Pankratov}}]{Shallcross2013}%
  \BibitemOpen
  \bibfield  {author} {\bibinfo {author} {\bibfnamefont {S.}~\bibnamefont
  {Shallcross}}, \bibinfo {author} {\bibfnamefont {S.}~\bibnamefont {Sharma}},
  \ and\ \bibinfo {author} {\bibfnamefont {O.}~\bibnamefont {Pankratov}},\
  }\href {\doibase 10.1103/PhysRevB.87.245403} {\bibfield  {journal} {\bibinfo
  {journal} {Phys. Rev. B - Condens. Matter Mater. Phys.}\ }\textbf {\bibinfo
  {volume} {87}},\ \bibinfo {pages} {245403} (\bibinfo {year}
  {2013})}\BibitemShut {NoStop}%
\bibitem [{\citenamefont {Sboychakov}\ \emph {et~al.}(2015)\citenamefont
  {Sboychakov}, \citenamefont {Rakhmanov}, \citenamefont {Rozhkov},\ and\
  \citenamefont {Nori}}]{Sboychakov2015}%
  \BibitemOpen
  \bibfield  {author} {\bibinfo {author} {\bibfnamefont {A.~O.}\ \bibnamefont
  {Sboychakov}}, \bibinfo {author} {\bibfnamefont {A.~L.}\ \bibnamefont
  {Rakhmanov}}, \bibinfo {author} {\bibfnamefont {A.~V.}\ \bibnamefont
  {Rozhkov}}, \ and\ \bibinfo {author} {\bibfnamefont {F.}~\bibnamefont
  {Nori}},\ }\href {\doibase 10.1103/PhysRevB.92.075402} {\bibfield  {journal}
  {\bibinfo  {journal} {Phys. Rev. B}\ }\textbf {\bibinfo {volume} {92}},\
  \bibinfo {pages} {075402} (\bibinfo {year} {2015})}\BibitemShut {NoStop}%
\bibitem [{\citenamefont {Cao}\ \emph {et~al.}(2016)\citenamefont {Cao},
  \citenamefont {Luo}, \citenamefont {Fatemi}, \citenamefont {Fang},
  \citenamefont {Sanchez-Yamagishi}, \citenamefont {Watanabe}, \citenamefont
  {Taniguchi}, \citenamefont {Kaxiras},\ and\ \citenamefont
  {Jarillo-Herrero}}]{jarilloHerrero2016}%
  \BibitemOpen
  \bibfield  {author} {\bibinfo {author} {\bibfnamefont {Y.}~\bibnamefont
  {Cao}}, \bibinfo {author} {\bibfnamefont {J.~Y.}\ \bibnamefont {Luo}},
  \bibinfo {author} {\bibfnamefont {V.}~\bibnamefont {Fatemi}}, \bibinfo
  {author} {\bibfnamefont {S.}~\bibnamefont {Fang}}, \bibinfo {author}
  {\bibfnamefont {J.~D.}\ \bibnamefont {Sanchez-Yamagishi}}, \bibinfo {author}
  {\bibfnamefont {K.}~\bibnamefont {Watanabe}}, \bibinfo {author}
  {\bibfnamefont {T.}~\bibnamefont {Taniguchi}}, \bibinfo {author}
  {\bibfnamefont {E.}~\bibnamefont {Kaxiras}}, \ and\ \bibinfo {author}
  {\bibfnamefont {P.}~\bibnamefont {Jarillo-Herrero}},\ }\href {\doibase
  10.1103/PhysRevLett.117.116804} {\bibfield  {journal} {\bibinfo  {journal}
  {Phys. Rev. Lett.}\ }\textbf {\bibinfo {volume} {117}},\ \bibinfo {pages}
  {116804} (\bibinfo {year} {2016})},\ \Eprint
  {http://arxiv.org/abs/1607.05147} {arXiv:1607.05147} \BibitemShut {NoStop}%
\bibitem [{\citenamefont {Kim}\ \emph {et~al.}(2016)\citenamefont {Kim},
  \citenamefont {Herlinger}, \citenamefont {Moon}, \citenamefont {Koshino},
  \citenamefont {Taniguchi}, \citenamefont {Watanabe},\ and\ \citenamefont
  {Smet}}]{Kim2016}%
  \BibitemOpen
  \bibfield  {author} {\bibinfo {author} {\bibfnamefont {Y.}~\bibnamefont
  {Kim}}, \bibinfo {author} {\bibfnamefont {P.}~\bibnamefont {Herlinger}},
  \bibinfo {author} {\bibfnamefont {P.}~\bibnamefont {Moon}}, \bibinfo {author}
  {\bibfnamefont {M.}~\bibnamefont {Koshino}}, \bibinfo {author} {\bibfnamefont
  {T.}~\bibnamefont {Taniguchi}}, \bibinfo {author} {\bibfnamefont
  {K.}~\bibnamefont {Watanabe}}, \ and\ \bibinfo {author} {\bibfnamefont
  {J.~H.}\ \bibnamefont {Smet}},\ }\href {\doibase
  10.1021/acs.nanolett.6b01906} {\bibfield  {journal} {\bibinfo  {journal}
  {Nano Lett.}\ }\textbf {\bibinfo {volume} {16}},\ \bibinfo {pages} {5053}
  (\bibinfo {year} {2016})}\BibitemShut {NoStop}%
\bibitem [{\citenamefont {Polshyn}\ \emph {et~al.}(2019)\citenamefont
  {Polshyn}, \citenamefont {Yankowitz}, \citenamefont {Chen}, \citenamefont
  {Zhang}, \citenamefont {Watanabe}, \citenamefont {Taniguchi}, \citenamefont
  {Dean},\ and\ \citenamefont {Young}}]{Polshyn2019}%
  \BibitemOpen
  \bibfield  {author} {\bibinfo {author} {\bibfnamefont {H.}~\bibnamefont
  {Polshyn}}, \bibinfo {author} {\bibfnamefont {M.}~\bibnamefont {Yankowitz}},
  \bibinfo {author} {\bibfnamefont {S.}~\bibnamefont {Chen}}, \bibinfo {author}
  {\bibfnamefont {Y.}~\bibnamefont {Zhang}}, \bibinfo {author} {\bibfnamefont
  {K.}~\bibnamefont {Watanabe}}, \bibinfo {author} {\bibfnamefont
  {T.}~\bibnamefont {Taniguchi}}, \bibinfo {author} {\bibfnamefont {C.~R.}\
  \bibnamefont {Dean}}, \ and\ \bibinfo {author} {\bibfnamefont {A.~F.}\
  \bibnamefont {Young}},\ }\href {http://arxiv.org/abs/1902.00763} {\
  (\bibinfo {year} {2019})},\ \Eprint {http://arxiv.org/abs/1902.00763}
  {arXiv:1902.00763} \BibitemShut {NoStop}%
\bibitem [{\citenamefont {Liu}\ \emph {et~al.}(2020)\citenamefont {Liu},
  \citenamefont {Wang}, \citenamefont {Watanabe}, \citenamefont {Taniguchi},
  \citenamefont {Vafek},\ and\ \citenamefont {Li}}]{Liu2020}%
  \BibitemOpen
  \bibfield  {author} {\bibinfo {author} {\bibfnamefont {X.}~\bibnamefont
  {Liu}}, \bibinfo {author} {\bibfnamefont {Z.}~\bibnamefont {Wang}}, \bibinfo
  {author} {\bibfnamefont {K.}~\bibnamefont {Watanabe}}, \bibinfo {author}
  {\bibfnamefont {T.}~\bibnamefont {Taniguchi}}, \bibinfo {author}
  {\bibfnamefont {O.}~\bibnamefont {Vafek}}, \ and\ \bibinfo {author}
  {\bibfnamefont {J.~I.~A.}\ \bibnamefont {Li}},\ }\href@noop {} {\  (\bibinfo
  {year} {2020})},\ \Eprint {http://arxiv.org/abs/2003.11072}
  {arXiv:2003.11072} \BibitemShut {NoStop}%
\bibitem [{\citenamefont {Yoo}\ \emph {et~al.}(2019)\citenamefont {Yoo},
  \citenamefont {Engelke}, \citenamefont {Carr}, \citenamefont {Fang},
  \citenamefont {Zhang}, \citenamefont {Cazeaux}, \citenamefont {Sung},
  \citenamefont {Hovden}, \citenamefont {Tsen}, \citenamefont {Taniguchi},
  \citenamefont {Watanabe}, \citenamefont {Yi}, \citenamefont {Kim},
  \citenamefont {Luskin}, \citenamefont {Tadmor}, \citenamefont {Kaxiras},\
  and\ \citenamefont {Kim}}]{Yoo2019}%
  \BibitemOpen
  \bibfield  {author} {\bibinfo {author} {\bibfnamefont {H.}~\bibnamefont
  {Yoo}}, \bibinfo {author} {\bibfnamefont {R.}~\bibnamefont {Engelke}},
  \bibinfo {author} {\bibfnamefont {S.}~\bibnamefont {Carr}}, \bibinfo {author}
  {\bibfnamefont {S.}~\bibnamefont {Fang}}, \bibinfo {author} {\bibfnamefont
  {K.}~\bibnamefont {Zhang}}, \bibinfo {author} {\bibfnamefont
  {P.}~\bibnamefont {Cazeaux}}, \bibinfo {author} {\bibfnamefont {S.~H.}\
  \bibnamefont {Sung}}, \bibinfo {author} {\bibfnamefont {R.}~\bibnamefont
  {Hovden}}, \bibinfo {author} {\bibfnamefont {A.~W.}\ \bibnamefont {Tsen}},
  \bibinfo {author} {\bibfnamefont {T.}~\bibnamefont {Taniguchi}}, \bibinfo
  {author} {\bibfnamefont {K.}~\bibnamefont {Watanabe}}, \bibinfo {author}
  {\bibfnamefont {G.-C.}\ \bibnamefont {Yi}}, \bibinfo {author} {\bibfnamefont
  {M.}~\bibnamefont {Kim}}, \bibinfo {author} {\bibfnamefont {M.}~\bibnamefont
  {Luskin}}, \bibinfo {author} {\bibfnamefont {E.~B.}\ \bibnamefont {Tadmor}},
  \bibinfo {author} {\bibfnamefont {E.}~\bibnamefont {Kaxiras}}, \ and\
  \bibinfo {author} {\bibfnamefont {P.}~\bibnamefont {Kim}},\ }\href {\doibase
  10.1038/s41563-019-0346-z} {\bibfield  {journal} {\bibinfo  {journal} {Nat.
  Mater.}\ }\textbf {\bibinfo {volume} {18}},\ \bibinfo {pages} {448} (\bibinfo
  {year} {2019})},\ \Eprint {http://arxiv.org/abs/1804.03806}
  {arXiv:1804.03806} \BibitemShut {NoStop}%
\bibitem [{\citenamefont {Andelkovi{\'{c}}}\ \emph {et~al.}(2018)\citenamefont
  {Andelkovi{\'{c}}}, \citenamefont {Covaci},\ and\ \citenamefont
  {Peeters}}]{Andelkovic2018}%
  \BibitemOpen
  \bibfield  {author} {\bibinfo {author} {\bibfnamefont {M.}~\bibnamefont
  {Andelkovi{\'{c}}}}, \bibinfo {author} {\bibfnamefont {L.}~\bibnamefont
  {Covaci}}, \ and\ \bibinfo {author} {\bibfnamefont {F.~M.}\ \bibnamefont
  {Peeters}},\ }\href {\doibase 10.1103/PhysRevMaterials.2.034004} {\bibfield
  {journal} {\bibinfo  {journal} {Phys. Rev. Mater.}\ }\textbf {\bibinfo
  {volume} {2}},\ \bibinfo {pages} {034004} (\bibinfo {year} {2018})},\ \Eprint
  {http://arxiv.org/abs/1705.05731} {arXiv:1705.05731} \BibitemShut {NoStop}%
\bibitem [{\citenamefont {Hwang}\ and\ \citenamefont
  {Sarma}(2019)}]{Hwang2019a}%
  \BibitemOpen
  \bibfield  {author} {\bibinfo {author} {\bibfnamefont {E.~H.}\ \bibnamefont
  {Hwang}}\ and\ \bibinfo {author} {\bibfnamefont {S.~D.}\ \bibnamefont
  {Sarma}},\ }\href {http://arxiv.org/abs/1907.02856} {\  (\bibinfo {year}
  {2019})},\ \Eprint {http://arxiv.org/abs/1907.02856} {arXiv:1907.02856}
  \BibitemShut {NoStop}%
\bibitem [{\citenamefont {Wu}\ \emph {et~al.}(2019{\natexlab{b}})\citenamefont
  {Wu}, \citenamefont {Hwang},\ and\ \citenamefont {{Das Sarma}}}]{Wu2019c}%
  \BibitemOpen
  \bibfield  {author} {\bibinfo {author} {\bibfnamefont {F.}~\bibnamefont
  {Wu}}, \bibinfo {author} {\bibfnamefont {E.}~\bibnamefont {Hwang}}, \ and\
  \bibinfo {author} {\bibfnamefont {S.}~\bibnamefont {{Das Sarma}}},\ }\href
  {\doibase 10.1103/PhysRevB.99.165112} {\bibfield  {journal} {\bibinfo
  {journal} {Phys. Rev. B}\ }\textbf {\bibinfo {volume} {99}},\ \bibinfo
  {pages} {165112} (\bibinfo {year} {2019}{\natexlab{b}})}\BibitemShut
  {NoStop}%
\bibitem [{\citenamefont {Pelc}\ \emph {et~al.}(2015)\citenamefont {Pelc},
  \citenamefont {Morell}, \citenamefont {Brey},\ and\ \citenamefont
  {Chico}}]{Pelc2015}%
  \BibitemOpen
  \bibfield  {author} {\bibinfo {author} {\bibfnamefont {M.}~\bibnamefont
  {Pelc}}, \bibinfo {author} {\bibfnamefont {E.~S.}\ \bibnamefont {Morell}},
  \bibinfo {author} {\bibfnamefont {L.}~\bibnamefont {Brey}}, \ and\ \bibinfo
  {author} {\bibfnamefont {L.}~\bibnamefont {Chico}},\ }\href {\doibase
  10.1021/acs.jpcc.5b00685} {\bibfield  {journal} {\bibinfo  {journal} {J.
  Phys. Chem. C}\ }\textbf {\bibinfo {volume} {119}},\ \bibinfo {pages} {10076}
  (\bibinfo {year} {2015})},\ \Eprint {http://arxiv.org/abs/1407.6594}
  {arXiv:1407.6594} \BibitemShut {NoStop}%
\bibitem [{\citenamefont {Bahamon}\ \emph {et~al.}(2019)\citenamefont
  {Bahamon}, \citenamefont {G{\'{o}}mez-Santos},\ and\ \citenamefont
  {Stauber}}]{Bahamon2019}%
  \BibitemOpen
  \bibfield  {author} {\bibinfo {author} {\bibfnamefont {D.~A.}\ \bibnamefont
  {Bahamon}}, \bibinfo {author} {\bibfnamefont {G.}~\bibnamefont
  {G{\'{o}}mez-Santos}}, \ and\ \bibinfo {author} {\bibfnamefont
  {T.}~\bibnamefont {Stauber}},\ }\href@noop {} {\  (\bibinfo {year} {2019})},\
  \Eprint {http://arxiv.org/abs/1909.09341} {arXiv:1909.09341} \BibitemShut
  {NoStop}%
\bibitem [{\citenamefont {Padhi}\ \emph {et~al.}(2020)\citenamefont {Padhi},
  \citenamefont {Tiwari}, \citenamefont {Neupert},\ and\ \citenamefont
  {Ryu}}]{Padhi2020}%
  \BibitemOpen
  \bibfield  {author} {\bibinfo {author} {\bibfnamefont {B.}~\bibnamefont
  {Padhi}}, \bibinfo {author} {\bibfnamefont {A.}~\bibnamefont {Tiwari}},
  \bibinfo {author} {\bibfnamefont {T.}~\bibnamefont {Neupert}}, \ and\
  \bibinfo {author} {\bibfnamefont {S.}~\bibnamefont {Ryu}},\ }\href
  {http://arxiv.org/abs/2005.02406} {\  (\bibinfo {year} {2020})},\ \Eprint
  {http://arxiv.org/abs/2005.02406} {arXiv:2005.02406} \BibitemShut {NoStop}%
\bibitem [{\citenamefont {Slater}\ and\ \citenamefont
  {Koster}(1954)}]{Slater1954}%
  \BibitemOpen
  \bibfield  {author} {\bibinfo {author} {\bibfnamefont {J.~C.}\ \bibnamefont
  {Slater}}\ and\ \bibinfo {author} {\bibfnamefont {G.~F.}\ \bibnamefont
  {Koster}},\ }\href {\doibase 10.1103/PhysRev.94.1498} {\bibfield  {journal}
  {\bibinfo  {journal} {Phys. Rev.}\ }\textbf {\bibinfo {volume} {94}},\
  \bibinfo {pages} {1498} (\bibinfo {year} {1954})},\ \Eprint
  {http://arxiv.org/abs/1506.08190} {arXiv:1506.08190} \BibitemShut {NoStop}%
\bibitem [{\citenamefont {{Trambly De Laissardi{\`{e}}re}}\ \emph
  {et~al.}(2012)\citenamefont {{Trambly De Laissardi{\`{e}}re}}, \citenamefont
  {Mayou},\ and\ \citenamefont {Magaud}}]{TramblyDeLaissardiere2012}%
  \BibitemOpen
  \bibfield  {author} {\bibinfo {author} {\bibfnamefont {G.}~\bibnamefont
  {{Trambly De Laissardi{\`{e}}re}}}, \bibinfo {author} {\bibfnamefont
  {D.}~\bibnamefont {Mayou}}, \ and\ \bibinfo {author} {\bibfnamefont
  {L.}~\bibnamefont {Magaud}},\ }\href {\doibase 10.1103/PhysRevB.86.125413}
  {\bibfield  {journal} {\bibinfo  {journal} {Phys. Rev. B - Condens. Matter
  Mater. Phys.}\ }\textbf {\bibinfo {volume} {86}},\ \bibinfo {pages} {125413}
  (\bibinfo {year} {2012})}\BibitemShut {NoStop}%
\bibitem [{\citenamefont {Groth}\ \emph {et~al.}(2014)\citenamefont {Groth},
  \citenamefont {Wimmer}, \citenamefont {Akhmerov},\ and\ \citenamefont
  {Waintal}}]{Groth2014}%
  \BibitemOpen
  \bibfield  {author} {\bibinfo {author} {\bibfnamefont {C.~W.}\ \bibnamefont
  {Groth}}, \bibinfo {author} {\bibfnamefont {M.}~\bibnamefont {Wimmer}},
  \bibinfo {author} {\bibfnamefont {A.~R.}\ \bibnamefont {Akhmerov}}, \ and\
  \bibinfo {author} {\bibfnamefont {X.}~\bibnamefont {Waintal}},\ }\href
  {\doibase 10.1088/1367-2630/16/6/063065} {\bibfield  {journal} {\bibinfo
  {journal} {New J. Phys.}\ }\textbf {\bibinfo {volume} {16}},\ \bibinfo
  {pages} {063065} (\bibinfo {year} {2014})},\ \Eprint
  {http://arxiv.org/abs/1309.2926} {arXiv:1309.2926} \BibitemShut {NoStop}%
\bibitem [{\citenamefont {Olyaei}\ \emph {et~al.}(2019)\citenamefont {Olyaei},
  \citenamefont {Ribeiro},\ and\ \citenamefont {Castro}}]{Olyaei2019}%
  \BibitemOpen
  \bibfield  {author} {\bibinfo {author} {\bibfnamefont {H.~Z.}\ \bibnamefont
  {Olyaei}}, \bibinfo {author} {\bibfnamefont {P.}~\bibnamefont {Ribeiro}}, \
  and\ \bibinfo {author} {\bibfnamefont {E.~V.}\ \bibnamefont {Castro}},\
  }\href {\doibase 10.1103/PhysRevB.99.205436} {\bibfield  {journal} {\bibinfo
  {journal} {Phys. Rev. B}\ }\textbf {\bibinfo {volume} {99}},\ \bibinfo
  {pages} {205436} (\bibinfo {year} {2019})},\ \Eprint
  {http://arxiv.org/abs/1810.11743} {arXiv:1810.11743} \BibitemShut {NoStop}%
\bibitem [{\citenamefont {{Su{\'{a}}rez Morell}}\ \emph
  {et~al.}(2011)\citenamefont {{Su{\'{a}}rez Morell}}, \citenamefont {Vargas},
  \citenamefont {Chico},\ and\ \citenamefont {Brey}}]{SuarezMorell2011}%
  \BibitemOpen
  \bibfield  {author} {\bibinfo {author} {\bibfnamefont {E.}~\bibnamefont
  {{Su{\'{a}}rez Morell}}}, \bibinfo {author} {\bibfnamefont {P.}~\bibnamefont
  {Vargas}}, \bibinfo {author} {\bibfnamefont {L.}~\bibnamefont {Chico}}, \
  and\ \bibinfo {author} {\bibfnamefont {L.}~\bibnamefont {Brey}},\ }\href
  {\doibase 10.1103/PhysRevB.84.195421} {\bibfield  {journal} {\bibinfo
  {journal} {Phys. Rev. B - Condens. Matter Mater. Phys.}\ }\textbf {\bibinfo
  {volume} {84}},\ \bibinfo {pages} {195421} (\bibinfo {year} {2011})},\
  \Eprint {http://arxiv.org/abs/1112.5467} {arXiv:1112.5467} \BibitemShut
  {NoStop}%
\bibitem [{\citenamefont {van~der Linden}\ \emph {et~al.}(2012)\citenamefont
  {van~der Linden}, \citenamefont {Doye},\ and\ \citenamefont
  {Louis}}]{VanderLinden2012}%
  \BibitemOpen
  \bibfield  {author} {\bibinfo {author} {\bibfnamefont {M.~N.}\ \bibnamefont
  {van~der Linden}}, \bibinfo {author} {\bibfnamefont {J.~P.~K.}\ \bibnamefont
  {Doye}}, \ and\ \bibinfo {author} {\bibfnamefont {A.~A.}\ \bibnamefont
  {Louis}},\ }\href {\doibase 10.1063/1.3679653} {\bibfield  {journal}
  {\bibinfo  {journal} {J. Chem. Phys.}\ }\textbf {\bibinfo {volume} {136}},\
  \bibinfo {pages} {54904} (\bibinfo {year} {2012})}\BibitemShut {NoStop}%
\bibitem [{\citenamefont {Moon}\ \emph {et~al.}(2019)\citenamefont {Moon},
  \citenamefont {Koshino},\ and\ \citenamefont {Son}}]{Moon2019}%
  \BibitemOpen
  \bibfield  {author} {\bibinfo {author} {\bibfnamefont {P.}~\bibnamefont
  {Moon}}, \bibinfo {author} {\bibfnamefont {M.}~\bibnamefont {Koshino}}, \
  and\ \bibinfo {author} {\bibfnamefont {Y.~W.}\ \bibnamefont {Son}},\ }\href
  {\doibase 10.1103/PhysRevB.99.165430} {\bibfield  {journal} {\bibinfo
  {journal} {Phys. Rev. B}\ }\textbf {\bibinfo {volume} {99}},\ \bibinfo
  {pages} {165430} (\bibinfo {year} {2019})}\BibitemShut {NoStop}%
\bibitem [{\citenamefont {Bistritzer}\ and\ \citenamefont
  {MacDonald}(2010)}]{bistritzer2010transport}%
  \BibitemOpen
  \bibfield  {author} {\bibinfo {author} {\bibfnamefont {R.}~\bibnamefont
  {Bistritzer}}\ and\ \bibinfo {author} {\bibfnamefont {A.~H.}\ \bibnamefont
  {MacDonald}},\ }\href {\doibase 10.1103/PhysRevB.81.245412} {\bibfield
  {journal} {\bibinfo  {journal} {Phys. Rev. B}\ }\textbf {\bibinfo {volume}
  {81}},\ \bibinfo {pages} {245412} (\bibinfo {year} {2010})},\ \Eprint
  {http://arxiv.org/abs/1002.2983} {arXiv:1002.2983} \BibitemShut {NoStop}%
\bibitem [{\citenamefont {Amorim}(2018)}]{Amorim2018}%
  \BibitemOpen
  \bibfield  {author} {\bibinfo {author} {\bibfnamefont {B.}~\bibnamefont
  {Amorim}},\ }\href {\doibase 10.1103/PhysRevB.97.165414} {\bibfield
  {journal} {\bibinfo  {journal} {Phys. Rev. B}\ }\textbf {\bibinfo {volume}
  {97}},\ \bibinfo {pages} {165414} (\bibinfo {year} {2018})},\ \Eprint
  {http://arxiv.org/abs/1711.02499} {arXiv:1711.02499} \BibitemShut {NoStop}%
\bibitem [{\citenamefont {Amorim}\ and\ \citenamefont
  {Castro}(2018)}]{Amorim2018b}%
  \BibitemOpen
  \bibfield  {author} {\bibinfo {author} {\bibfnamefont {B.}~\bibnamefont
  {Amorim}}\ and\ \bibinfo {author} {\bibfnamefont {E.~V.}\ \bibnamefont
  {Castro}},\ }\href {http://arxiv.org/abs/1807.11909} {\  (\bibinfo {year}
  {2018})},\ \Eprint {http://arxiv.org/abs/1807.11909} {arXiv:1807.11909}
  \BibitemShut {NoStop}%
\end{thebibliography}%

\end{document}